\documentclass[12pt]{article}
\usepackage{amsfonts,epsfig,latexsym,amscd,multicol,theorem}

\newcommand{\sech}{{\rm sech}\, }
\newcommand{\kd}{\ker{\Dd^{\phi_{\mu}}}}
\newcommand{\vm}{V_{\mu}}
\newcommand{\sm}{s_{\mu}}
\newcommand{\cm}{c_{\mu}}
\newcommand{\km}{K_{\mu}}
\newcommand{\intinf}{\int_{-\infty}^{\infty}{ ds\ }}
\newcommand{\hp}{\hess_{\phi}}
\newcommand{\hpn}{\hess_{\phi_n}}
\newcommand{\vh}{\langle V,V \rangle_{H^1_\phi}}
\newcommand{\vhn}{\langle V,V \rangle_{H^1_{\phi_n}}}
\newcommand{\vhnn}{\langle V_n,V_n \rangle_{H^1_{\phi_n}}}
\newcommand{\nv}{\nabla V}
\newcommand{\nnv}{\nabla_n V}

\newcommand{\R}{{\mathbb{R}}}

\newcommand{\C}{{\mathbb{C}}}
\newcommand{\I}{{\mathbb{I}}}

\newcommand{\CP}{{\mathbb{C}}{{P}}}
\newcommand{\RP}{{\mathbb{R}{{P}}}}
\newcommand{\beq}{\begin{equation}}
\newcommand{\eeq}{\end{equation}}
\newcommand{\bea}{\begin{eqnarray}}
\newcommand{\eea}{\end{eqnarray}}
\newcommand{\ra}{\rightarrow}
\newcommand{\rhu}{\rightharpoonup}
\newcommand{\hra}{\hookrightarrow}

\newcommand{\cd}{\partial}
\newcommand{\wt}{\widetilde}
\newcommand{\wh}{\widehat}
\newcommand{\M}{{\sf M}}
\newcommand{\msn}{{\sf M}_n}
\newcommand{\mst}{{\sf M}_2}
\newcommand{\hess}{{\sf Hess}}
\newcommand{\hol}{{\sf Hol}}
\newcommand{\rat}{{\sf Rat}}
\newcommand{\half}{\frac{1}{2}}
\newcommand{\su}{{\mathfrak{su}}}
\newcommand{\so}{{\mathfrak{so}}}
\newcommand{\jac}{{\mathfrak J}}
\newcommand{\Dd}{{\mathfrak D}}
\newcommand{\ric}{{\mathfrak R}}
\newcommand{\en}{{\cal E}}
\newcommand{\Vv}{{\cal V}}
\newcommand{\Uu}{{\cal U}}
\newcommand{\ph}{{\varphi}}

\newcommand{\ol}{\overline}

\newcommand{\sigvec}{\mbox{\boldmath{$\sigma$}}}
\newcommand{\lamvec}{\mbox{\boldmath{$\lambda$}}}
\newcommand{\tauvec}{\mbox{\boldmath{$\tau$}}}

\newcommand{\zerovec}{\mbox{\boldmath{$0$}}}

\newcommand{\tr}{{\rm tr}}
\newcommand{\cosec}{{\rm cosec}\, }
\newcommand{\id}{{\rm Id}}

\newcommand{\eps}{\epsilon}

\newcommand{\be}{\begin{equation}}
\newcommand{\ee}{\end{equation}}

\theoremstyle{plain}
\newtheorem{thm}{Theorem}
\newtheorem{lemma}[thm]{Lemma}
\newtheorem{prop}[thm]{Proposition}
\newtheorem{cor}[thm]{Corollary}
\newtheorem{conj}[thm]{Conjecture}
{\theorembodyfont{\rmfamily}

}

\newcommand{\news}{\setcounter{equation}{0}}

\begin{document}

\title{The geodesic approximation for lump dynamics
and coercivity of the Hessian for harmonic maps}
\author{M. Haskins\thanks{E-mail: {\tt mhaskin@math.jhu.edu}} \\
Department of Mathematics, Johns Hopkins University\\
Baltimore, MD 21218, U.S.A.\\ \\
J.M. Speight\thanks{E-mail: {\tt j.m.speight@leeds.ac.uk}}\\
Department of Pure Mathematics, University of Leeds\\
Leeds LS2 9JT, England}

\date{}
\maketitle

\begin{abstract}
The most fruitful approach to studying low energy soliton dynamics in field
theories of Bogomol'nyi type is the geodesic approximation of Manton. In the
case of vortices and monopoles, Stuart has obtained rigorous estimates
of the errors in this approximation, and hence proved that it is valid in the
low speed regime. His method employs energy estimates which rely on a key
coercivity property of the Hessian of the energy functional of the theory
under consideration. In this paper we prove an analogous coercivity property
for the Hessian of the energy functional of a general sigma model with
compact K\"ahler domain and target. We go on to prove a continuity property
for
our result, and show that, for the $\CP^1$ model on $S^2$, the Hessian
fails to be globally coercive in the degree $1$ sector. We present numerical
evidence which suggests that the Hessian is globally coercive in a certain
equivariance class of the degree $n$ sector for $n\geq 2$. We also prove
that, within the geodesic approximation, a single $\CP^1$ lump moving on
$S^2$ does not generically travel on a great circle.

\end{abstract}

\maketitle

\section{Introduction}
\label{sec:intro}
\news

Many field theories arising naturally in theoretical high energy physics may
be said to be of Bogomol'nyi type. For such theories there is a topological
lower bound on the energy of field configurations, and this bound is attained
only by solutions of a first order ``self-duality'' equation, the so-called
solitons of the theory. The solitons are stable by virtue of their
energy-minimizing property, and are generically spatially localized lumps of
energy with strongly particle-like characteristics. When static they exert
no net force on one another, so the structure of the space of static
multisoliton solutions is rather rich. Examples within the context of gauge
theory are given by the Yang-Mills-Higgs and abelian-Higgs models, whose
solitons are called monopoles and vortices respectively. In both these cases,
the static models are very well understood and the structure of $\msn$,
the
moduli space of static $n$-soliton solutions, is known in great detail.
For monopoles, in particular, the static system is integrable and there are
several constructions which generate exact solutions of various degrees of
explicitness. Once one introduces time dependence, however, things get much
more difficult. No Bogomol'nyi type field theory (indeed, no Lorentz
invariant
field theory) in more than $(1+1)$ dimensions is integrable, and the
construction of nontrivial exact time-dependent solutions seems
impossible.

How then is one to understand the dynamics of moving solitons in
these models?
The most fruitful approach
has turned out to be the geodesic approximation of Manton
\cite{man1}. Here one argues
on physical grounds that the solution of
any initial value problem in the $n$-soliton
sector whose initial field is a static solution, and whose initial kinetic
energy is small, should be forced to stay close to $\msn$
 by energy conservation. Manton suggested that the
dynamics should be well approximated by a solution of the reduced
variational problem where the field configuration is {\em constrained} to
lie on $\msn$ at all times. This reduced dynamics turns out to be geodesic
flow on $\msn$ with respect to a natural metric called the $L^2$ metric, so
one has the appealing suggestion that low energy soliton dynamics in
Bogomol'nyi type theories may be understood by studying the Riemannian
geometry of their moduli spaces. This is still a highly nontrivial problem,
and it was some time before the $L^2$ metrics on the 2-monopole and 2-vortex
moduli spaces were well understood
\cite{atihit,sam}. Since first being proposed, the method
of Manton has been extended to deal with dynamical issues other than
classical multisoliton scattering. Quantum soliton states
\cite{gibman} are thought to be
well approximated by eigenstates of the Laplacian on $\msn$, for example,
and the thermodynamics of soliton gases \cite{mannas}
has been analyzed in terms of the
geometry of $\msn$.

Not all Bogomol'nyi type field theories are gauge theories. Another class is
given by nonlinear sigma models with K\"ahler target space, for example the
$\CP^N$ models. These models are nonlinear in the most fundamental way: the
field takes values in a space with no linear structure. However they have
many features in common with the gauge theories mentioned above. There is
again a topological lower bound on energy, attained only by solutions of
a first order partial differential equation, namely the $\pm$-holomorphic
maps -- the Cauchy-Riemann conditions play the role of the self-duality
equation. The static model is integrable and the structure of $\msn$ is
again well understood. The solitons in this case are usually called
``lumps.''

Given the similarities between lumps and their gauge theoretic
counterparts, it was natural for Ward to suggest, in the specific context of
the $\CP^1$ model on the plane, that the geodesic approximation should be
applicable to classical lump dynamics too \cite{war}. A detailed numerical
analysis of 2-lump scattering within the geodesic approximation followed
\cite{lee}, as well as generalizations to the $\CP^N$ models \cite{stozak}.
One technical problem encountered in all these studies is that $\CP^N$ lumps
have $L^2$-infinite zero modes on $\C$, so the $L^2$ metric is only
well-defined on the leaves of a foliation of $\msn$, rather than $\msn$
itself. One interprets this physically as saying that certain parameters in
the static $n$-lump solution are frozen to constant values by infinite
inertia. For example, the width of a single $\CP^1$ lump is a free
parameter, but is frozen in this fashion. Unfortunately, this freezing
appears to be an artifact of the approximation -- numerical solutions suggest
that a single lump may expand or contract according to the genuine field
dynamics. So the geodesic approximation is rather pathological for these
models. It is therefore interesting to consider situations where physical
space is compact, since the $L^2$ metric is then guaranteed to be
well-defined.
In particular, $\CP^1$ lumps on $S^2$ and $T^2$ have been studied, and quite
a lot is known about the corresponding $L^2$ geometries
\cite{spe1,spe2,spe3}.

The question remains, of course, whether geodesic flow in $\msn$ really does
closely approximate low-energy $n$-lump dynamics in these theories. For
2-vortex and 2-monopole dynamics, rigorous results supporting the geodesic
approximation have been proved by Stuart \cite{stu1,stu2}. He has shown that
the solution with initial data $\phi(0)\in\mst$, $\dot{\phi}(0)\in
T_{\phi(0)}\mst$ of order $\epsilon$ (where $\epsilon>0$ is small), stays
pointwise close (order $\epsilon^2$)
to its corresponding geodesic in $\mst$ for a time of order $\epsilon^{-1}$.
The key idea in Stuart's analysis is to separate the dynamics into slow
and fast time-varying modes by means of the following projection: the
true solution $\phi(t)$ is projected onto $\mst$ to obtain a slow
trajectory $\wt{\phi}$
\beq\label{1}
\phi(t)=\wt{\phi}(\epsilon t)+\epsilon^2 V(t)
\eeq
and a fast varying ``error'' $V(t)$, the projection being chosen so that
$V(t)$ is always $L^2$ orthogonal to $T_{\wt{\phi}(\eps t)}\mst$. Stuart
goes on to prove that an appropriate Sobolev norm of $V(t)$ remains bounded
for times of order $\epsilon^{-1}$ by bounding this norm in terms of
$\hess_{\wt{\phi}}$, the quadratic form associated with the second variation
of the potential energy functional of the model, which is slowly varying by
virtue of energy conservation. It follows that, as expected, the true
solution remains close to $\mst$.
That the projected trajectory $\wt{\phi}$
remains close to geodesic is proved as a separate step.

The whole analysis relies on one's ability to control Sobolev norms of the
error in terms of the Hessian. More precisely, the following
coercivity property of $\hess$ is crucial:
\vspace{0.25cm}

\noindent
{\it There exists a constant $\tau>0$ such that for all $\wt{\phi}\in\mst$
and all $V\in H^1$ with $V\perp_{L^2} T_{\wt{\phi}}\mst$,
$\hess_{\wt{\phi}}(V,V)\geq\tau ||V||_{H^1}^2$.}
\vspace{0.25cm}

\noindent
This turns out to be slightly easier to prove for vortices than monopoles
since the $L^2$ spectrum of small oscillations about a static vortex has a
mass gap due to the Higgs mechanism. No such gap occurs for monopoles.

How much of this framework carries over to sigma models? Since the target
space has no linear structure, equation (\ref{1}) makes no sense as it
stands, and must be replaced. We suggest here that the correct replacement
is
\beq
\phi(t)=\exp_{\wt{\phi}(\epsilon t)}\epsilon^2 V(t)
\eeq
where $\exp:TN\ra N$ is the exponential map on the target space $N$. Once
again, we should choose $V(t)$ always to be $L^2$ orthogonal to
$T_{\wt{\phi}(\epsilon t)}\msn$. In this context, we will prove an analogue
of Stuart's coercivity lemma for any holomorphic map $\wt{\phi}$ between
compact K\"ahler manifolds, namely, there exists $\tau(\wt{\phi})>0$ such
that
for all
$V\in H^1$, $V\perp_{L^2}
T_{\wt{\phi}}\msn$,
\beq
\hess_{\wt{\phi}}(V,V)\geq \tau(\wt{\phi})
||V||_{H^1}^2.
\eeq
The important difference from Stuart's result is that the constant $\tau$
depends on $\wt{\phi}$, that is, varies with position in $\msn$. We may
define the optimal constant
\beq
\tau(\wt{\phi})=\inf\left\{
\frac{\hess_{\wt{\phi}}(V,V)}{||V||_{H^1}^2}: V\neq 0,
V\perp_{L^2}T_{\wt{\phi}}\msn\right\}>0.
\eeq
We prove a result (Theorem \ref{taucty}) which gives sufficient conditions
that $\tau$ depend continuously on $\wt{\phi}$, and verify that those
conditions hold
in the main examples of interest to us. It turns out that
$\msn$ for sigma models is generically non-compact, so continuity of
$\tau(\wt{\phi})$ does not guarantee a global bound.
 In fact, we will show by means of an explicit counterexample, that
$\tau(\wt{\phi})>0$ is {\em not} necessarily bounded away from zero. The
counterexample occurs in the simplest nontrivial case, namely the one-lump
moduli space of the $\CP^1$ model on $S^2$, but we believe it is indicative
of a generic phenomenon for the $\CP^N$ models on any compact Riemann
surface. Roughly speaking, $\tau(\wt{\phi})\ra 0$ as the lump $\wt{\phi}$
shrinks to zero size. One would expect this to happen quite generically in
$\msn$, wherever a family of holomorphic maps degenerates so that a single
isolated lump collapses. This suggests we are never likely to have a global
bounding constant $\tau$ on $\msn$  as obtained by Stuart for
vortices and monopoles. Consequently, the best result one could hope for
from an analysis of this type for lumps is that the geodesic approximation
is good for a time of order $T\epsilon^{-1}$, where $T$ is some increasing
function of $\inf_t\tau(\wt{\phi}(\epsilon t))$. In this case, the
approximation would work well (in the usual sense)
so long as the projected geodesic stays away from
$\cd_\infty\msn$, the
boundary at infinity of $\msn$,
 where $\tau\ra 0$. In particular it seems very unlikely that geodesic
flow provides a good approximation to the process of single lump collapse
itself. Two independent numerical studies of singularity formation in the
$\CP^1$ model on the plane
 support this pessimistic assessment \cite{bizchmtab,linsad}.

If we impose extra symmetry on our system, in other words, restrict attention
to an admissible equivariance class,
then the projected
map $\wt{\phi}$ is confined to a totally geodesic submanifold $\M_n^{eq}$ of
$\M_n$,
and the error $V$ is confined
to an infinite dimensional subspace $H^1_{eq}$ of $H^1(\wt{\phi}^*TN)$.
We may define an equivariant version $\tau^{eq}$ of $\tau$ by taking
the infimum only over sections $V\in H^1_{eq}$. Clearly $\tau^{eq}(\wt{\phi})
\geq\tau(\wt{\phi})>0$. For a certain equivariance class for
$n$-lump dynamics in the $\CP^1$ model on $S^2$, one can prove that
$\tau^{eq}$ is
continuous on $\M_n^{eq}$, which is again noncompact.
We will present numerical evidence
 that $\tau^{eq}(\wt{\phi})$ {\em is} globally bounded
away from zero for $n\geq 2$ in this class.
The point is that only {\em coincident}
$n$-lump collapse can
occur within this particular
equivariance class, and the problem of vanishing $\tau$
does not appear to happen for such collapse. It is possible, therefore,
that the geodesic approximation does give a good model of equivariant
multilump collapse, though one should be cautious: there is ample scope for
other aspects of Stuart's method to break down as singularities form. There
are some grounds for optimism. A recent careful numerical study of lump
collapse by Linhart and Sadun found that single lump collapse on the plane
differed significantly from that predicted by a truncated geodesic
approximation, while coincident 2-lump collapse did not \cite{linsad}.

The rest of this paper is structured as follows. In section \ref{sec:cp1}
we introduce the nonlinear sigma models of interest and briefly review
some of their standard properties. We focus in particular on the $\CP^1$
model
on $S^2$, describing what is known about the $L^2$ geometry on $\M_n$ in this
case, especially for $n=1$. We present a new result on generic behaviour of
the geodesic flow in $\M_1$ which implies that (within the geodesic
approximation) a single lump generically does {\em not} travel along great
circles on $S^2$. Precise restrictions on the set of non-generic initial
data are given. In section \ref{coerce} we formulate and prove
the main result of the
paper, that the Hessian for these models is coercive on $H^1$, in the sense
described above. In section \ref{cty} we establish a simple sufficient
condition for $\tau(\phi)$ to depend continuously on $\wt{\phi}\in\M_n$.
In section \ref{rat1} we show that this condition is met in the case of
$\M_1$ for the $\CP^1$ model on $S^2$. We go on to prove that $\tau\ra 0$
as $\wt{\phi}$ approaches $\cd_\infty\M_1$, so global coercivity fails in
this
case. Finally, in section \ref{ratn} we consider the case of the $n$-lump
sector of the $\CP^1$ model on $S^2$ within a given equivariance class,
showing that $\tau^{eq}$ is continuous. We conjecture that $\hess^{eq}$ is
globally coercive for $n\geq 2$ and present some numerical results in support
of this conjecture.

\section{The $\CP^1$ model on $S^2$}
\label{sec:cp1}\news

For our purposes, a nonlinear sigma model consists of a single field
$\phi:\R\times M\ra N$, where $(M,g)$ and $(N,h)$ are compact K\"ahler
manifolds, $M$ represents physical space, $\R\times M$ is spacetime, equipped
with the Lorentzian pseudometric $\eta=dt^2-g$ and $N$ is the target space.
Solutions of the model are local extremals of the action
\beq
S[\phi]=\frac{1}{2}\int_{\R\times M}
\sum_\alpha\frac{|d\phi E_\alpha|^2}{|E_\alpha|^2}=\int_\R\left(\half\int_M
|\dot{\phi}|^2-E[\phi]\right)
\eeq
where $E_0=\cd/\cd t,E_1,\cdots,E_m$ is an orthonormal basis of vector
fields on $\R\times M$ and $E[\phi]$ is the harmonic map energy functional
for maps $M\ra N$. Such solutions are called wave maps in the geometric
analysis literature, by analogy with harmonic maps
\cite{shastr}. Indeed, static wave
maps are precisely harmonic maps and hence have been the focus of intense
study. In particular, Lichnerowicz showed that if a homotopy class $[\phi]$
contains $\pm$-holomorphic representatives, then these minimize energy within
that class
\cite{lic}. So the moduli space of interest within a given class is
$\M_{[\phi]}=\hol_{[\phi]}(M,N)$. For the sake of generality, we will state
and prove the coercivity lemma for the Hessian in this general context.

It is important when using the geodesic approximation  to know that
$\M_{[\phi]}$
is a finite dimensional smooth manifold. This is not always true in the
general case of holomorphic maps between K\"ahler manifolds.
We discuss this question in more
detail in section \ref{coerce},
briefly summarizing some results in the harmonic maps and algebraic geometry
literature which allow us to identify classes of sigma model whose
moduli spaces are smooth manifolds.
For the moment we note that
 the particular case we have most directly in mind,
namely $M=N=S^2$, the $\CP^1$ model on the two-sphere, certainly does have
this property.

We now consider the case $M=N=S^2$ in more detail.
Each homotopy class $[\phi]\in\pi_2(S^2)$ is labelled by the degree of
$\phi$, an integer $n$, which without loss of generality we may assume is
non-negative. The degree $n$ is interpreted physically as the net lump
number of the configuration $\phi$. The space $\M_n$ of degree $n$
holomorphic maps $S^2\ra S^2$ is easily constructed explicitly. Choosing
stereographic coordinates $z,W\in \C$ on both $M$ and $N$, such a map is
rational of degree $n$,
\beq
\phi:z\mapsto W=\frac{a_1z^n+\cdots+a_{n+1}}{a_{n+2}z^n+\cdots+a_{2n+2}}
\eeq
where $a_i$ are $2n+2$ complex constants, $a_1$ and $a_{n+2}$ do not both
vanish, and the numerator and denominator have no common roots. Since
$(a_1,\ldots,a_{2n+2})$ and $(\xi a_1,\ldots,\xi a_{2n+2})$ give the same
map for all $\xi\in\C\backslash\{0\}$, we have a natural identification of
$\M_n$ with a dense open subset of $\CP^{2n+1}$, whence it inherits a natural
topology and complex structure.

The metric of interest $\gamma$, does not descend from the inclusion
$\M_n\subset\CP^{2n+1}$ however. To define it, one must think of a tangent
vector $X\in T_\phi\M_n$ as a zero mode of the Cauchy-Riemann equations for
maps $M\ra N$ at the map $\phi$. Such a zero mode is a smooth section of
$\phi^*TN$, the pullback of the tangent bundle of $N$ by the map $\phi$,
that is, a rule which assigns to each $p\in M$ a vector
$V(p)\in T_{\phi(p)}N$.
We may define the $L^2$ inner product between any pair of sufficiently
regular sections $X,Y$ of $\phi^*TN$ by taking their fibrewise inner product
in $T_{\phi(p)}N$, then integrating over $M$,
\beq
\langle X,Y\rangle_{L^2} =\int_M h(X,Y).
\eeq
The $L^2$ metric on $M_n$ is simply the restriction of $\langle\cdot,\cdot
\rangle_{L^2}$
to zero modes. In more concrete terms, one can in principle compute
explicit formulae for $\gamma$ by choosing local coordinates $q^i$ on
$\M_n$ (for example, the real and imaginary parts of $a_i/a_1$, on the
chart where $a_1\neq 0$) and expressing the map $\phi(\{q^i\})$ as an
explicit
function $W(z,\{q^i\})$, so that
\beq
\gamma=\sum_{ij}\gamma_{ij}dq^idq^j,\quad
\gamma_{ij}=\int_\C\frac{dzd\ol{z}}{(1+|z|^2)^2}\frac{1}{(1+|W|^2)^2}
\frac{\cd W}{\cd q^i}\frac{\ol{\cd W}}{\cd q^j}.
\eeq
In practice, of course, the integrals involved are almost always intractable.

Certainly, the $L^2$ metric is a natural way of geometrizing $\M_n$. More
importantly, it is the Riemannian metric descending from the restriction of
the kinetic energy functional $\half\int_M|\dot{\phi}|^2$ to $\M_n$, and
hence
the metric whose geodesics are thought to model slow lump dynamics. We shall
briefly review what is known about the Riemannian manifold $(\M_n,\gamma)$,
and prove a new result about the generic behaviour of geodesics in $(\M_1,
\gamma)$.

First, $(\M_n,\gamma)$ is manifestly Hermitian, and is in fact K\"ahler. This
was long suspected, owing to a rather general formal argument of Ruback
\cite{rub}, and
has recently been proved rigorously \cite{spe3}. It is also known that
$(\M_n,\gamma)$ is geodesically incomplete \cite{sadspe}. For odd $n$,
$(\M_n,\gamma)$ contains a totally geodesic Lagrangian submanifold
naturally identified with the moduli space of static $\RP^2$ $n$-lumps on
$\RP^2$; for $n\geq 3$ this submanifold is also incomplete.

There is an
isometric action of $G=SO(3)\times SO(3)$ on $\M_n$, induced by the natural
$SO(3)$ actions on the domain and target spheres, which on $\M_1$ has
cohomogeneity 1 (generic $G$ orbits have codimension 1), and in fact, almost
completely determines $\gamma$. Consequently, an explicit formula for
$\gamma$ is known in this case, and the geometry is particularly well
understood. For $n=1$, the no common roots condition on the rational map
$W(z)=(a_1z+a_2)/(a_3z+a_4)$ is $a_1a_4-a_2a_3\neq0$, so we may identify each
map with a projective equivalence class $[L]$
of $GL(2,\C)$ matrices. Hence
$\M_1\cong PL(2,\C)$. By identifying $S^2$ with the unit sphere in
$\R^3\cong \su(2)$ in the usual way, we may identify the $SO(3)$ action on
$S^2$ with the adjoint $SU(2)$ action, so that $\gamma$ regarded as a metric
on $PL(2,\C)$ is invariant under the left and right $PU(2)$ actions:
\beq
([U_1],[U_2]):[L]\mapsto [U_1LU_2].
\eeq
Now every $[L]\in PL(2,\C)$ has a unique polar decomposition
\beq
[L]=[U(\sqrt{1+\lambda^2}\I_2+\lamvec\cdot\tauvec)]
\eeq
where $([U],\lamvec)\in PU(2)\times\R^3$,
$\lambda=|\lamvec|$, and $\tau_1,\tau_2,\tau_3$ are the
Pauli spin matrices. Hence $\M_1\cong PU(2)\times\R^3\cong SO(3)\times\R^3$.
Physically, the lump corresponding to $([U],\lamvec)$ has maximum
energy density
 at $-\lamvec/\lambda\in S^2$, sharpness proportional to $\lambda$ and
internal orientation $[U]$. The $\lambda=0$ lumps have uniform energy
density. The $G$ action in this coordinate system is
\beq
([U_1],[U_2]):([U],\lamvec)\mapsto ([U_1UU_2],Ad_{U_2}\lamvec)
\eeq
where again we have used $\R^3\cong\su(2)$ to identify the fundamental
$SO(3)$
action on $\R^3$ with the adjoint $SU(2)$ action on $\su(2)$. From this we
see that the $G$-orbits are level sets of $\lambda$, generically
diffeomorphic
to $SO(3)\times S^2$ (when $\lambda>0$), the only exception being
$\lambda=0$,
which is diffeomorphic to $SO(3)$.

In \cite{spe3} it was proved that every $G$ invariant K\"ahler metric on
$\M_1$ may be written
\beq
\gamma=A_1\, d\lamvec\cdot d\lamvec+A_2(\lamvec\cdot d\lamvec)^2+
A_3\, \sigvec\cdot\sigvec+A_4(\lamvec\cdot\sigvec)^2+A_1\lamvec\cdot
(\sigvec\times d\lamvec)
\eeq
where $A_1,\ldots,A_4$ are smooth functions of $\lambda$ only, all determined
from the single function $A_1=A(\lambda)$ by the relations
\beq
A_2=\frac{A(\lambda)}{1+\lambda^2}+\frac{A'(\lambda)}{\lambda},
\quad A_3=\frac{1}{4}(1+2\lambda^2)A(\lambda),
\quad A_4=\frac{1}{4\lambda}(1+\lambda^2)A'(\lambda).
\eeq
Here $\sigma_1,\sigma_2,\sigma_3$ are the left invariant one forms on
$SO(3)$ dual to the basis $\{\frac{i}{2}\tau_a:a=1,2,3\}$ for $\su(2)\cong
\so(3)$.  For the $L^2$ metric, one finds that
\beq\label{Adef}
A=\frac{4\pi\mu[\mu^4-4\mu^2\log\mu-1]}{(\mu^2-1)^3},
\quad \mu=(\sqrt{1+\lambda^2}+\lambda)^2.
\eeq
It follows from these formulae that $(\M_1,\gamma)$ has finite volume and
diameter, is Ricci positive and has unbounded scalar and holomorphic
sectional curvatures. Examining the large $\lambda$ behaviour of $\gamma$,
one finds that $\cd_\infty\M_1$, the boundary at infinity of $\M_1$ is
$S^2\times S^2$. This is natural in two ways: a point in $\cd_\infty\M_1$
should be thought of as a collapsed lump whose width has shrunk to zero.
Such a lump is specified by a pair of points $p,p'$ in $S^2$ because
every point except one, $p$, in the domain gets mapped to the same point
$p'$ in the codomain, while $p$ gets mapped to the antipodal point $-p'$.
Second, the complex codimension 1 algebraic variety $a_1a_4-a_2a_3=0$
complementary to $\M_1$ in $\CP^3$ is biholomorphic to $\CP^1\times \CP^1$,
being the image of the holomorphic embedding
\beq
([x_1,x_2],[y_1,y_2])\mapsto [x_1y_2,x_1y_1,x_2y_2,x_2y_1].
\eeq
Since ${\rm diam}(\M_1,\gamma)<\infty$, $\cd_\infty\M_1$ lies at finite
proper distance, so geodesics may reach it in finite time, the origin of the
incompleteness already noted. Given that $\cd_\infty\M_1$ has (real)
codimension
2, however, one would expect geodesics to miss $\cd_\infty\M_1$ generically.
More precisely, one would expect the subset of $T\M_1$ consisting of initial
data of geodesics which {\em do} escape to infinity to have zero measure
with respect to the natural measure inherited from $\gamma$.

Geodesic flow in $(\M_1,\gamma)$ was studied in detail in \cite{spe1}. It
turns out to be surprisingly complicated given the homogeneity and isotropy
of the domain of the sigma model, $S^2$. Geodesics were found for which the
lump spins internally and oscillates between antipodal points. Other
geodesics were found where the lump travels along a great circle in
$S^2$, its speed and shape undergoing complicated periodic oscillations.
Geodesics do exist for which the lump simply traverses a great circle at
constant speed and shape, but the initial data to generate such motion must
be chosen very carefully. Nevertheless, all the geodesics found in
\cite{spe1} confined the lump's position $-\lamvec/\lambda$ to  some
great circle for all time. More precisely, they all confined $\lamvec(t)$ to
some plane through the origin. The geodesics themselves were obtained by
reducing the geodesic problem to  low dimensional totally geodesic
submanifolds, the fixed point sets of discrete isometries. The question
arises, then, whether this ``planar property'' of the geodesic flow
is an artifact of the exceptional symmetries enjoyed by these geodesics,
or is a general feature of the dynamics.
We shall now prove that the former is the case.

By the $G$ invariance of the metric, it suffices to determine which initial
velocities $u\in T_{x(0)}H$
 at the point $x(0)=
([\I_2],(0,0,\lambda)$ tangent to the hypersurface
$H=\{([U],\lamvec):\lambda_2=0\}$ generate geodesics which remain in $H$.
We shall call such an initial velocity $u$ a ``good'' vector.
Clearly the set of good initial velocities is conical
by time-scaling
invariance of geodesic flow. By this, we mean that if $u\in T_{x(0)}H$ is
good, so is every vector $\xi u\in T_{x(0)}H$, $\xi\in[0,\infty)$, on the
ray containing $u$. The question is then whether the {\em link} of this
 cone of good velocity vectors, that is, its intersection with the unit
4-sphere in $T_{x(0)}H$, has non-vanishing measure in $S^4$.

\begin{thm} Let $u\in T_{x(0)}H$ generate a geodesic through $x(0)$ which
remains in $H$. Then $u$ lies in a codimension 1 cone in $T_{x(0)}H$.
For generic $x(0)\in H$, the link of this cone
is the suspension of a two-torus. There is at most a nowhere dense subset
of $H$ on which the link of the cone is $S^3\cup S^3$, two three-spheres
intersecting in an equatorial two-sphere.
\end{thm}

\noindent
{\it Proof:} Let $N$ be a nonvanishing (but not necessarily unit)
normal on $H$. Then if
$x(t)$ remains in $H$, $\gamma(\dot{x}(t),N(x(t)))=0$ for all $t$.
Differentiating this at $t=0$ and using the fact that $\dot{x}(t)$ is
parallel for a geodesic, one finds that
$\gamma(\dot{x}(0),\nabla_{\dot{x}(0)}
N)=0$, where $\nabla$ is the Levi-Civita connexion.
Hence, $u=\dot{x}(0)$ must lie in the null space of the symmetric
bilinear form
\beq
B(u,v)=\gamma(u,\nabla_vN),
\eeq
that is $B(u,u)=0$. Clearly this null space is conical.
We seek to understand the link of this cone.
In this case ($H=\{\lambda_2=0\}$), we may choose
\beq
N=A_3\frac{\cd\, }{\cd\lambda_2}+\frac{1}{2}A_1(-\lambda_3\theta_1+
\lambda_1\theta_3)
\eeq
as our normal field, where $\theta_a$ are the left-invariant vector fields
on $SO(3)$ dual to $\sigma_a$.
We may compute $B(u,v)$ by extending $u,v$ to vector
fields $U,V$ on $\M_1$, then using the usual formula for $\nabla$
and symmetry of $B$ \cite{wil}, to yield
\beq
B(u,v)=\frac{1}{2}
\left.N[\gamma(U,V)]\right|_{x(0)}+\frac{1}{2}\gamma(\left.[U,N]
\right|_{x(0)},v)+\frac{1}{2}\gamma(\left.[V,N]\right|_{x(0)},u).
\eeq
A straightforward but lengthy calculation then shows that
\beq
B=f_1(\lambda)d\lambda_1\sigma_3+f_2(\lambda)d\lambda_3\sigma_1+
f_3(\lambda)\sigma_2\sigma_3
\eeq
where
\beq
f_1=\frac{1}{8}(1+\lambda^2)(2A+\lambda A'),\quad
f_2=-\frac{1}{4}(1+2\lambda^2)A(A+\lambda A'),\quad
f_3=\frac{1}{16}(1+\lambda^2)AA',
\eeq
at the specific point $x(0)=([\I_2],(0,0,\lambda))$. By computing
eigenvalues, one sees that, with respect to some orthonormal basis for
$T_{x(0)}H$,
\beq
B(u,u)=f_2(\lambda)(u_2^2-u_3^2)+\sqrt{f_1(\lambda)^2+f_3(\lambda)^2}(u_4^2
-u_5^2).
\eeq
Note that
\beq
\frac{8f_1(\lambda)}{1+\lambda^2}=2A+\lambda A'>
\frac{2\lambda^1+1}{\lambda^2+1}A+\lambda A'=
\gamma(\frac{\cd\, }{\cd\lambda_3},\frac{\cd\, }{\cd\lambda_3})>0
\eeq
so $f_1(\lambda)$ never vanishes. Consider the set $L=f_2^{-1}(0)\subset
[0,\infty)$. If $L$ were dense at $\lambda_0$, then by continuity of
$A+\lambda A'$, there would be an open interval containing $\lambda_0$ such
that $A=-\log\lambda+\mbox{constant}$. Given the formula (\ref{Adef}) $A$
clearly does not coincide with $-\log\lambda+\mbox{constant}$ on any
interval,
so no such $\lambda_0$ exists. Hence the set $L$ is nowhere dense.

Let $\lambda\notin L$. By means of
a $\lambda$ dependent rescaling of the basis, we see that the null space is
the locus of the equation
\beq
\label{cone}
\tilde{u}_2^2+\tilde{u}_4^2=\tilde{u}_3^2+\tilde{u}_5^2.
\eeq
Clearly $\tilde{u}_1$ may take any value, while
$(\tilde{u}_2,\tilde{u}_3,\tilde{u}_4,\tilde{u}_5)$ lies on a cone in
$\R^4$ whose link is a two-torus. Alternatively, we may think of the
unit sphere in $\R^5$ as the suspension of the unit sphere in $\R^4$ along
the $\tilde{u}_1$ direction. Then the intersection of the link of the
null space with each 3 sphere of constant $\tilde{u}_1\in[-1,1]$ is a
scaled two-torus. Assembling the slices together we see that the whole null
space has a link which is topologically the suspension of a two-torus
\cite{hat}. Either
way of viewing the null set, it is clearly a codimension 1 cone in
$T_{x(0)}H$
as claimed.

It remains to consider the non-generic case, $\lambda\in L$, for which
$f_2(\lambda)=0$. In this case, since $f_1(\lambda)\neq 0$,
 the cone consists of all vectors for which
$u_4=\pm u_5$. The link is the intersection of $S^4$ with the union of the
two orthogonal hyperplanes $(u_4\pm u_5)=0$, which is manifestly a union of
two three-spheres intersecting in an equatorial two sphere ($u_4=u_5=0$).
\hfill$\Box$

\vspace{0.5cm}
We should point out that $u\in{\rm null}\, B$ is a necessary but not
sufficient
condition for the geodesic with initial velocity $u$ to stay in $H$. It is
not certain that the cone of good velocities is all of ${\rm null}\,
 B$,
therefore.
Given the explicit formula for $A$, one would expect to be able to
improve the characterization of the non-generic subset of $H$
(on which ${\rm null}\, B$ has link $S^3\cup S^3$) from nowhere dense to
discrete or finite. Indeed, one may check graphically that $f_2(\lambda)=0$
has only one solution ($\lambda=0.881$ to 3 decimal places), so
the nongeneric set consists of a single $G$ orbit, $S^1\times SO(3)$, in
$H$.

\begin{cor} Generically geodesics in $(\M_1,\gamma)$ do not confine
$\lamvec(t)
$ to a plane through $\zerovec$.
Consequently, single lumps generically do not stay on
great circles while moving on $S^2$.
\end{cor}

Of course, this corollary refers to the lump dynamics within the geodesic
approximation. The question remains: does this give a good model of the full
field dynamics?

\section{Coercivity of the Hessian}
\label{coerce}
\news

We wish to consider the wave map problem for maps $(\R\times M,dt^2-g)\ra
(N,h)$, where $(M,g),(N,h)$ are compact K\"ahler, and the initial data have
a certain Sobolev regularity and are close to holomorphic, in a sense to be
made precise. The eventual aim is to prove that such wave maps stay close
to $\hol_{[\phi]}(M,N)$ and that their closest trajectory in
$\hol_{[\phi]}(M,N)$ is close to a geodesic. In this section we will prove
the main analytic ingredient needed for such a programme, along the lines
of Stuart's work on vortices.

The first thing to note is that the Cauchy problem for such wave maps with
initial data $(\phi_0,\dot{\phi}_0)\in H^k\oplus H^{k-1}$ on the time slice
$\{0\}\times M$ is well posed, that is, has a unique solution in the
same Sobolev space, with $t\mapsto (\phi(t),\dot{\phi}(t))$ continuous,
at least on some open time interval, $t\in (-\eps,\eps),$ by
work of Choquet-Bruhat \cite{cho}. Here $H^k$ denotes the space of
maps $M\ra\R^p$ which are $H^k$ in the usual sense ($\phi, \nabla\phi,\ldots,
\nabla^k\phi$ are all $L^2$) and which take values on $N\subset\R^p$
(isometrically embedded in $\R^p$) almost everywhere.
For our purposes, it is convenient to use an alternative intrinsic
definition of $H^k(M,N)$, which is only well-defined
for $k> \frac{1}{2} \textrm{dim\,}M$. With this restriction
on $k$, $H^k(M,N)$ naturally has the structure of a Hilbert manifold.
Below we give a brief treatment of both the intrinsic definition
and the Hilbert manifold structure of $H^k(M,N)$. For a complete treatment
we refer the reader to Palais \cite{Palais}.

A map $\phi:M \ra N$ is said to belong to $H^k(M,N)$ if for any
$p\in M$ and any chart $(U,\Phi)$ containing $p$, and any chart
$(V,\Psi)$ containing $\phi(p)$, the map $\Psi \circ \phi \circ
\Phi^{-1}: \Phi(U) \ra \R^n$ belongs to $H^k(\Phi(U),\R^n)$. For this
notion to be well-defined we need to ensure that
composition by $C^\infty$ diffeomorphisms on the left and on the right
takes an $H^k$ map to an $H^k$ map. For composition on the
right this is true without any restriction on $k$. However, for
left composition the same result only holds if we assume
$k> \frac{1}{2} \textrm{dim\,}M$ (for a discussion of
both points see B.1.7 and B.1.9, p182 \cite{mcd-salamon}).
From now on we make a standing assumption that $k >
\frac{1}{2}\textrm{dim\,}M$, so that the intrinsic definition of $H^k(M,N)$
given above makes sense.

We now exhibit the structure of an infinite
dimensional Hilbert manifold on $H^k(M,N)$.
Standard facts on embeddings of Sobolev spaces and the density of
smooth maps in Sobolev spaces prove the following:
There are continuous inclusions $C^\infty (M,N) \hookrightarrow H^k(M,N)\hookrightarrow C^0(M,N)$ and
the first inclusion has dense image. From the latter fact it follows that it is sufficient
to exhibit charts for $H^k(M,N)$ around only the $C^\infty$ maps.

Given $\phi \in C^\infty(M,N)$ the pullback bundle $\phi^*TN$
(i.e. the vector bundle over $M$ whose fibre at $m$ is the vector space $T_{\phi(m)}N$)
comes equipped with a natural inner product $h\circ (\phi \times \phi)$ and compatible connexion $\nabla^{\phi}$,
the pullback of the Levi-Civita connexion on $TN$. For any vector bundle $E$ over compact $M$
equipped with an inner product $<\  ,\  >$ and compatible connection $\nabla$ there is a natural $H^k$ inner product
on smooth sections of $E$:
\beq
<V,W>_{H^k}=\int_M <V,W> +\int_M <\nabla V, \nabla W> +\cdots+
\int_M <\nabla^k V, \nabla^kW>.
\eeq
$H^k(E)$ is then defined as the set of finite $H^k$ norm elements of the
completion of $C^\infty(E)$ with respect to the norm $||\, . \,||_{H^k}$.
For $M$ compact, this definition of $H^k(E)$ is equivalent
to the following alternative definition: given any choice
of local coordinates on $M$ and associated bundle trivializations for $E$,
a section belongs to $H^k(E)$ if it is represented by locally
$H^k$ functions in these trivializations. The difference
is that the connexion-dependent definition gives a preferred
inner product on $H^k(E)$, i.e. $(H^k(E),||\, .\, ||_{H^k})$ is naturally
a Hilbert space.

The point is that $H^k(\phi^*TN)$ is the local model space for the
Hilbert manifold $H^k(M,N)$. For each $\phi\in C^\infty(M,N)$, there is
a map $\exp_{\phi}: H^k(\phi^*TN)\ra H^k(M,N)$ given by
$V(p) \mapsto \exp_{\phi(p)}{V(p)}$,
where $\exp:TN\ra N$ is the exponential map on $(N,h)$.
It can be shown that $\exp_{\phi}$ maps a neighbourhood of $0$ in the Hilbert space
$H^k(\phi^*TN)$ bijectively to a neighbourhood of $\phi$ in $H^k(M,N)$.
Hence for each $\phi \in C^\infty(M,N)$ there exists some $\eps>0$,
so that one can define a chart $(U_{\phi},\exp_{\phi}^{-1})$ based at $\phi$ where
$$U_{\phi}=\{\exp_{\phi}(V): V\in H^k(\phi^*TN), \  ||V||_{H^k}<\eps\}.$$
The local homeomorphism
$U_{\phi}\ra B_\eps(0)\subset H^k(\phi^*TN)$ is simply given by $\exp_{\phi}^{-1}$. So we identify $H^k$ maps close to $\phi$
with $H^k$ sections of $\phi^*TN$ by deforming $\phi$: the deformed map
$\exp_{\phi}(V)$ maps each $p\in M$ to the point in $N$ reached by travelling for
unit time along the geodesic with initial data $(\phi(p),V(p))$.

It can be shown that for any two maps $\phi_1$, $\phi_2 \in C^\infty(M,N)$
the change of charts map
$\exp^{-1}_{\phi_2} \circ \exp_{\phi_1}: \exp_{\phi_1}^{-1}(U_{\phi_1} \cap U_{\phi_2}) \ra
\exp_{\phi_2}^{-1}(U_{\phi_1} \cap U_{\phi_2})$
is a diffeomorphism between open sets in the Hilbert spaces $H^k(\phi_1^*TN)$ and $H^k(\phi_2^*TN)$.
It follows that the collection of charts $\{(U_{\phi},\exp^{-1}_\phi) |\  \phi \in C^{\infty}(M,N)\}$
defines a differentiable structure on $H^k(M,N)$ with local model a Hilbert space, i.e. $H^k(M,N)$
is a Hilbert manifold. In fact, the differentiable structure can be shown
to be independent of the metric $h$ on $N$ used to define $\exp_{\phi}$.

\medskip

Given
\begin{itemize}
\item a holomorphic map
$\wt{\phi}_0\in\hol(M,N),$ and
\item sections $V_0,X_0,Y_0\in H^k(\wt{\phi}_0^*TN)$, such that
\item
$V_0,Y_0$ are $L^2$ orthogonal to $T_{\wt{\phi}_0}\hol(M,N)$, and
\item $X_0\in
T_{\wt{\phi}_0}\hol(M,N)$
\end{itemize}
and $\eps>0$ small, the initial value problem with
initial data
\beq
\phi_0=\exp_{\wt{\phi}_0}\eps^2V_0\, ,\quad
\dot{\phi}_0=\eps X_0+\eps^3 Y_0
\eeq
has a unique solution in $H^k$. The idea is to decompose this solution into
$\wt{\phi}(t)\in\hol(M,N)$ and $V(t)\in H^k(\wt{\phi}(t)^*TN)\cap
T_{\wt{\phi}(t)}\hol(M,N)^{\perp_{L^2}}$ by
\beq
\phi(t)=\exp_{\wt{\phi}(t)}\eps^2V(t)
\eeq
and then show that $||V||_{H^k}$ remains bounded for times of order
$\eps^{-1}$. The starting point is to show that $||V||_{H^1}$ is controlled
by
$\hess_{\wt{\phi}}(V,V)$. From now on, all quantities will be considered at
a fixed time, and we will denote the holomorphic base map of our local
chart in $H^k$ by $\phi$ rather than $\wt{\phi}$, to simplify notation.

$\hess_\phi(X,Y)$ is the second variation of the harmonic map energy
$E$ at the holomorphic (hence harmonic) map $\phi$. Precisely, given a
two-parameter variation $\phi_{s,t}$ of the map $\phi$ through smooth maps,
with $d\phi_{0,0}\cd/
\cd s=X,d\phi_{0,0}\cd/\cd t=Y\in \Gamma(\phi^*TN)$,
\beq
\hess_\phi(X,Y)=\left.\frac{\cd^2 E[\phi_{s,t}]}{\cd s\cd t}\right|_{s=t=0}.
\eeq
There are two useful explicit formulae for $\hess$
\cite{ura}. The first uses only
compactness of $M$, not the K\"ahler property. To write it down we must
introduce two new objects. Let $E_i,i=1,\ldots,m$ be a local frame of
smooth orthonormal vector fields on $M$. Then the rough Laplacian on
sections of $\phi^*TN$ is the second order linear elliptic differential
operator
\beq\label{roughlap}
\Delta^\phi V=-\tr(\nabla^\phi\nabla^\phi V)
=-\sum_{i=1}^m(\nabla^\phi\nabla^\phi V)(E_i,E_i).
\eeq
Like the usual Laplacian (on functions or forms) $\Delta^\phi$ is a
positive self-adjoint operator. Positivity follows from the identity
\beq\label{pos}
\int_M h(V,\Delta^\phi V)=\half\int_M\sum_i h(\nabla_{E_i}^\phi V,
\nabla_{E_i}^\phi V).
\eeq
We may define a (fibrewise linear) bundle map $\ric^\phi$ on $\phi^*TN$ by
\beq
\label{ric}
\ric^\phi V=\sum_{i=1}^m R^N(V,d\phi\, E_i)d\phi\, E_i
\eeq
where $R^N$ is the curvature tensor on $N$. Given these, the
Hessian is
\beq\label{hess1}
\hess_\phi(X,Y)=\int_M h(X,\jac^\phi Y)=
\langle X,\jac^\phi \rangle_{L^2},\quad
\jac^\phi=\Delta^\phi-\ric^\phi.
\eeq
The operator $\jac^\phi$, called the Jacobi operator,
 is itself second order, linear, elliptic and self-adjoint.
 It follows immediately from
this formula that every harmonic map into a manifold of nonpositive sectional
curvature is weakly stable (meaning $\hess_\phi(V,V)\geq 0$).

In the case of interest to us, namely $M,N$ compact K\"ahler,
and $\phi$ holomorphic, one can obtain
a more useful and rather simpler formula for $\hess$. First one defines
the Urakawa connexion on $\phi^*TN$,
\beq
(\Dd^\phi V)(X):=\nabla^\phi_{J^MX}V-J^N\nabla^\phi_XV,
\eeq
$J^M,J^N$ being the almost complex structures on $M$ and $N$ \cite{ura}. Then
\beq\label{*}
\hess_{\phi}(V,V)=\frac{1}{2}||\Dd^\phi V||_{L^2}^2.
\eeq
Weak stability of holomorphic maps follows immediately from (\ref{*}).

$\Dd^\phi V=0$ should be thought of as the linearized Bogomol'nyi equation,
so every $V\in\ker\Dd^\phi=\ker\jac^\phi$ is a zero mode of the
holomorphic map $\phi$. For the geodesic approximation to make sense, the
space of holomorphic maps close to $\phi$ must be a smooth manifold whose
tangent space at $\phi$ equals $\ker\jac^\phi$.
Sections of $\phi^*TN$ in the kernel of $\jac^\phi$
are called Jacobi fields along $\phi$.
A Jacobi field is said to be integrable if it may
be generated by a variation of $\phi$ through harmonic maps, and the map
itself is said to be Jacobi integrable if all its Jacobi fields have this
property.
A fundamental theorem of Adams and Simon
\cite{adasim} states that a harmonic map $\phi$
between real analytic manifolds
is Jacobi integrable if and only if the space of  harmonic maps
$C^{2,\alpha}$ close to $\phi$ is a smooth manifold with tangent space
$\ker\jac^\phi$ at $\phi$. Similar results hold in suitable Sobolev spaces
of maps also.
 In the case where $M,N$ are K\"ahler
(hence real analytic),
Lichnerowicz showed that all harmonic deformations of a holomorphic
map $\phi$ are holomorphic \cite{lic}, so the space of interest to us, the
space of
{\em holomorphic} maps close to $\phi$, is a smooth manifold with
tangent space $\ker\jac^\phi$ at $\phi$, if and only if $\phi$ is Jacobi
integrable. Jacobi integrability of harmonic maps is an active
 field of research whose current state is summarized in
 \cite{lemwoo}. Particularly relevant to the present paper is a theorem of
Wood and Lemaire which states that every holomorphic map $S^2\ra \CP^N$ is
Jacobi integrable. So the moduli space of holomorphic maps for the $\CP^N$
model on $S^2$ is smooth with tangent space $\ker\jac^\phi$, as we
require. A similar result holds for degree $n$ holomorphic maps
$\Sigma\ra\CP^1$, $\Sigma$ being a compact Riemann surface of genus $g$,
provided $n>2g-2$, by a standard application of the Riemann-Roch
theorem. So the $n$-lump moduli space of the $\CP^1$ model on $\Sigma$ also
has the required property, with some low degree exceptions.

We may now state and prove our main result.

\begin{thm}[Coercivity of the Hessian]\label{coercivity} Let $\phi:M\ra N$
be a holomorphic
map between compact K\"ahler manifolds and $\hess_\phi$ be the Hessian of
the harmonic map energy functional at $\phi$. Then there exists a constant
$\tau(\phi)>0$ such that for all $V\in H^1(\phi^*TN)$ with
$\langle V,\ker\Dd^\phi\rangle_{L^2}=0$,
$$
\hess_\phi(V,V)\geq \tau(\phi)||V||_{H^1}^2.
$$
\end{thm}

\noindent
{\it Proof:} Both $H^1$ and $L^2$ are Hilbert spaces. We will use
$\ra$ and $\rhu$ to denote strong and weak convergence respectively, the
space concerned being explicitly specified.
 Define the subset $S=\{V\in H^1:||V||_{H^1}^2=1, V\perp_{L^2}
\ker\Dd^\phi\}$ and the quantity
\beq
\tau(\phi)=\inf_{V\in S}\hess_\phi(V,V)=\inf_{V\in S}\half||\Dd^\phi V
||_{L^2}^2
\geq 0.
\eeq
We claim that $\tau(\phi)\neq 0$. Assume this is false. Then there exists
a sequence $V_i\in S$ such that $\Dd^\phi V_i\stackrel{L^2}{\ra} 0$.
We will repeatedly extract (nested) subsequences from $V_i$, which we will
always denote by the same symbol, $V_i$.
 Now
$V_i$ is bounded in $H^1$ so by the Alaoglu theorem, there exists a
subsequence $V_i\stackrel{H^1}{\rhu} V$, to some weak limit $V$. Since
$\Dd^\phi:H^1\ra L^2$ is a bounded linear map, it is continuous with
respect to the weak (and strong) topologies on $H^1,L^2$. It follows that
$\Dd^\phi V_i\stackrel{L^2}{\rhu} \Dd^\phi V$. But $\Dd^\phi V_i
\stackrel{L^2}{\ra}
0\, \Rightarrow\,  \Dd^\phi V_i\stackrel{L^2}{\rhu}0\, \Rightarrow\,
\Dd^\phi V=0$ by uniqueness of weak limits. Hence $V_i\stackrel{H^1}{\rhu}
V\in\ker\Dd^\phi$.

Now the inclusion $\iota:H^1\hra L^2$ is compact
by Rellich's lemma, so the bounded set $\{V_i\}\subset H^1$ is compact
in $L^2$. Hence, any sequence in $\{V_i\}$, for example $V_i$ itself, has
 a subsequence which is strongly convergent in $L^2$. Once again,
denoting this
subsequence by $V_i$, we have
$V_i\stackrel{L^2}{\ra}\tilde{V}$. But then $V_i\stackrel{L^2}{\rhu}
\tilde{V}$
so
\beq
0=\langle V_i,\ker\Dd^\phi\rangle_{L^2}\ra\langle \tilde{V},\ker\Dd^\phi
\rangle_{L^2}.
\eeq
Hence $V_i\stackrel{L^2}{\ra}\tilde{V}\in(\ker\Dd^\phi)^{\perp_{L^2}}$.

But $\iota:H^1\hra L^2$ is continuous, so $V_i\stackrel{H^1}{\rhu} V\,
\Rightarrow\, \iota V_i\stackrel{L^2}{\rhu}\iota V$, and so $\tilde{V}=V$
by uniqueness of weak limits. So $V\in\ker\Dd^\phi$ and $V\in(\ker\Dd^\phi)^{
\perp_{L^2}}$, and hence $V=0$, so $V_i\stackrel{L^2}{\ra}0$.
Then, by (\ref{pos}) and (\ref{hess1}),
\bea
1&=& ||V_i||_{H^1}^2=||V_i||_{L^2}^2+\sum_\alpha ||\nabla_{E_\alpha}^\phi
V_i||_{L^2}^2\nonumber \\
&=&2\hess_\phi(V_i,V_i)+||V_i||_{L^2}^2+\sum_\alpha
2\langle R^N(d\phi E_\alpha, V_i)d\phi E_\alpha,V_i\rangle_{L^2}\nonumber\\
&\leq&2\hess_\phi(V_i,V_i)+||V_i||_{L^2}^2+C_\phi||V_i||_{L^2}^2
\label{**}
\eea
where $C_\phi>0$ is a constant,
by compactness of $N$ and the tensorial property of $\ric^N$. Taking limits
of both sides of (\ref{**})
 and using $\tau(\phi)=0$ and $V_i\stackrel{L^2}{\ra}
0$, one sees that $1\leq 0$, a contradiction.\hfill
$\Box$

\vspace{0.5cm}

We remark that the $L^2$ version of this result, that there exists a constant
$\tau(\phi)>0$ such that $\hess_\phi(V,V)\geq \tau(\phi)||V||_{L^2}^2$ for
all $V\perp_{L^2}\ker\jac^\phi$, is much easier to prove. Since $\jac^\phi$
is elliptic, self adjoint and positive definite (for weakly stable $\phi$),
and $M$ is compact, we know immediately that the spectrum of $\jac^\phi$
is discrete and, normal to its kernel, bounded away from $0$. The result
immediately follows. In fact the optimal constant $\tau_{L^2}(\phi)$ in
this case is just the lowest nonvanishing eigenvalue of $\jac^\phi$. Hence
the optimal bound is attained in this case, by any eigensection with this
eigenvalue.
We remark also that the spectrum of $\jac^\phi$ is
of independent physical interest since it gives the semiclassical
meson spectrum of the sigma model in the topological sector $[\phi]$.

\section{Continuity of $\tau(\phi)$}\label{cty}
\news

Let $\phi_n:M \ra N$ be a sequence of smooth holomorphic maps between
compact
K\"ahler manifolds, converging
in $C^1$ to a smooth holomorphic map $\phi:M \ra N$. From Theorem
\ref{coercivity} we have for each $n$ a positive constant $\tau(\phi_n)$
and the positive constant $\tau(\phi)$, which give lower bounds for the
ratio $\hess(V,V)/\langle V,V \rangle_{H_1}$ for any
$H^1$ section of the pullback bundle $L^2$ orthogonal to all Jacobi fields.
In this section we will establish
conditions sufficient to guarantee that $\lim_{n\ra \infty}{\tau(\phi_n)}=
\tau(\phi)$. In subsequent sections
we will show that these conditions are met in the cases of interest to us.

In order to compare various quantities (especially the Hessian and $H^1$
norm)
 at different maps, it is convenient to make
various identifications so that we can treat all geometric quantities and
operators as being defined on the fixed bundle $E=\phi^*TN$.
Since $\phi_n \ra \phi$, for sufficiently large $n$ each pullback bundle
$E_n=\phi_n^*TN$ is topologically equivalent to
$E$. However, since each bundle $E_n$ comes naturally equipped with both an
inner product and a compatible connexion (both of which
occur in the Hessian and the $H^1$-norm of a section) we would also like to
transfer these geometric structures to the bundle $E$.

Again since $\phi_n \ra \phi$, for each $x\in M$ (and for each sufficiently
large $n$) there is a unique minimizing geodesic joining $\phi_n(x)$
to $\phi(x)$. By parallel transporting vectors along this unique geodesic we
construct a canonical isometry between the fibres of $E$ and the fibres
of $E_n$ at each point $x\in M$, and hence a natural $L^2$ isometry between
$E$ and $E_n$. Using this isometry we can interpret the
natural connexion on each $E_n$ as a connexion on $E$, which we shall write
$\nabla_n$. Using the connexion $\nabla_n$ and the (fixed) $L^2$ metric
on $E$ we can now interpret each of the Jacobi operators $\jac^{\phi_n}$
(and
hence also the associated symmetric bilinear form $\hess_{\phi_n}$) as an
operator on sections of the fixed bundle $E$.
Similarly, for each $n$ we get an $H^1_{\phi_n}$ norm on sections of $E$, by
using the connection $\nabla_n$ and the fixed $L^2$ metric.
Since the difference of any two connexions on $E$ is tensorial, it is clear
that for each $n$ (sufficiently large) the $H^1_{\phi_n}$ norm is equivalent
to the $H^1_{\phi}$ norm
on sections of $E$.

Suppose now that the difference between the connexions $\nabla_n$ and
$\nabla$ tends to zero in the the following pointwise sense
\beq
|\nabla V - \nabla_n V| \le a_n |V| \label{ptwise}
\eeq
where each $a_n$ is a positive number, $a_n\ra 0$ as $n\ra \infty$, and
$|.|$ refers to the natural pointwise norms on the bundles
$E\otimes T^*M$ and $E$, and $V$ is a smooth section of $E$.

\begin{lemma}
\label{hesscty}
Suppose that (\ref{ptwise}) holds, then the following inequalities also hold:

(i) $|\  \parallel\nabla V\parallel^2 - \parallel\nabla_n V\parallel^2 \ |
\le b_n \parallel V\parallel_1^2$

(ii) $|\ \parallel V\parallel^2_1 - \parallel V\parallel^2_{H^1_{\phi_n}}\ |
\le b_n \parallel V\parallel_1^2$

(iii) $|\ \hess_{\phi}(V,V) - \hess_{\phi_n}(V,V)\ | \le c_n \parallel V
\parallel_1^2$

\noindent
where $b_n$, $c_n$ are positive numbers which tend to zero as $n\ra \infty$,
and
$\parallel.\parallel$ is the $L^2$ norm on $\Gamma(E)$,
$\parallel.\parallel_1$ is the $H^1$ norm on $\Gamma(E)$
defined using the connection $\nabla$ and $\parallel.
\parallel_{H^1_{\phi_n}}$
 is the $H^1$ norm on $\Gamma(E)$
defined using the connection $\nabla_n$.
\end{lemma}

\noindent {\it Proof:} (i) Elementary manipulations involving the triangle
inequality, inequality (\ref{ptwise}) and the Cauchy-Schwarz inequality
yield the following chain of inequalities
\bea
\left | \int_M{ \left(|\nv|^2 -| \nnv|^2\right) } \right| &\le&
\int_M{ \left|
 |\nv|^2 -|\nnv|^2\right| } \nonumber \\
&\le & \int_M{\left| \  |\nv|-|\nnv| \ \right| \ .  \left|\  |\nv| +
 |\nnv| \ \right| } \nonumber \\
&\le& \int_M{\left| \  |\nv|-|\nnv| \ \right| \ .  \left( \left|\  2|\nv| +
 |\nv-\nnv| \ \right| \right) } \nonumber \\
&\le& 2a_n\int_M{|V| \ |\nv|\ } + a_n^2 \int_M{|V|^2 } \nonumber \\
&\le& 2a_n \parallel\ |V| \ \parallel. \parallel\ |\nv|\ \parallel + a_n^2
\parallel V\parallel^2 \nonumber \\
&\le& ( 2a_n + a_n^2) \ \parallel V \parallel_1^2\ . \nonumber
\eea
Hence the result follows with $b_n = a_n(2+a_n)$.\\
(ii) This is immediate from the definition of the $H^1$ norm and part (i).\\
(iii) The definition of $\ric^{\phi}$ (see equation \ref{ric}), together
with
the compactness of $M$ and the fact that
$\phi_n \ra \phi$ in $C^1$ implies that
\beq
\label{riccty}
\lim_{n\ra \infty}{\parallel \ric^{\phi}-\ric^{\phi_n}\parallel}=0.
\eeq
Using part (i) and the definition of $\hess$ we have
\bea
\left| \hp(V,V) - \hpn(V,V) \right| &=& \left| \frac{1}{2}\parallel \nv
\parallel^2 - \frac{1}{2}\parallel \nnv\parallel^2
-\langle \ric^{\phi}V,V\rangle + \langle \ric^{\phi_n}V,V\rangle \right|
\nonumber \\
&\le& \frac{1}{2}
\left| \parallel \nv \parallel^2 - \parallel \nnv\parallel^2 \right|
+ \left| \langle \ric^{\phi}V,V\rangle - \langle \ric^{\phi_n}V,V\rangle
\right| \nonumber\\
&\le& b_n \parallel V \parallel^2_1 + \parallel (\ric^{\phi}-\ric^{\phi_n})V
 \parallel \ \parallel V \parallel \nonumber \\
&\le& b_n \parallel V \parallel^2_1 + \parallel \ric^{\phi}-\ric^{\phi_n}
\parallel \ \parallel V \parallel^2. \nonumber
\eea
The result now follows from (\ref{riccty}). \hfill $\Box$\\

We will also need the following simple proposition on the continuity of
$L^2$ orthogonal projection operators.

\begin{prop}
Let $H^k(E)$ denote the $H^k$ sections of $E$. Let $V_n$ be a sequence of
finite
dimensional vector subspaces of $H^k(E)$ of constant dimension $p$
which converge to another $p$-dimensional subspace $V$ of $H^k(E)$,
in the following sense: there exists an $L^2$-orthonormal basis $e_1^n,
\ldots ,
e_p^n$ of $V_n$ and an $L^2$-orthonormal basis $e_1, \ldots , e_p$ of $V$ so
that for
each $i=1, \ldots p$, $e_i^n \ra e_i$ in $H^k$ as $n\ra \infty$.
Denote by $P_n$ and $P$ the $L^2$ orthogonal projections onto
$V_n^{\perp}$ and $V^{\perp}$ respectively. Then $P_n \ra P$ in
the operator norm on $B(H^k(E),H^k(E))$, the bounded linear maps between
$H^k(E)$ and itself.
\end{prop}

\noindent {\it Proof:}
In this proof, norms and inner products without subscripts will refer to
$L^2$ norms and inner products, while $H^k$
norms and inner products will be referred to with the subscript $k$.
Since $P_n v = v - \sum_i{\langle v, e_i^n \rangle e_i^n}$ and $Pv = v -
\sum_i{\langle v, e_i \rangle e_i}$
we have
\bea
||P_n v - Pv ||_k &=& ||\sum_i{\langle v,e_i^n \rangle e^n_i} -
\sum_i{\langle v,e_i \rangle e_i} ||_k \le
\sum_i{||\langle v,e_i^n \rangle e^n_i - {\langle v,e_i \rangle e_i} ||_k
\nonumber} \\
&\le& \sum_i{||\langle v,e_i^n \rangle e^n_i - \langle v,e_i^n \rangle e_i +
\langle v,e_i^n \rangle e_i -\langle v,e_i \rangle e_i ||_k} \nonumber \\
&\le& \sum_i{|\langle v,e_i^n \rangle|\  || e_i^n-e_i||_k + ||e_i||_k \
|\langle v,e_i^n-e_i \rangle|} \nonumber \\
&\le& ||v||\ \left(\sum_i{||e_i^n||\ ||e_i^n-e_i||_k + ||e_i||_k\
||e_i^n-e_i||}\right) \nonumber \\
&\le& ||v|| \left(\sum_i{(1+ ||e_i||_k)\ ||e_i^n - e_i||_k}\right).
\eea
So for all $v\neq 0 \in H^k$ we have,
\beq
\frac{||(P_n-P)v||_k}{||v||_k} \le \frac{||(P_n-P)v||_k}{||v||} \le
\left(\sum_i{(1+ ||e_i||_k)\ ||e_i^n - e_i||_k}\right). \label{pnminusp}
\eeq
Since by assumption $e_i^n \ra e_i$ strongly in $H^k$ for each $i$,
(\ref{pnminusp}) implies that
\beq
\lim_{n\ra \infty}{\sup_{v\neq0}{\frac{||(P_n-P)v||_k}{||v||_k}}}=0
\eeq
as required.
\hfill $\Box$\\

We now state the main result of this section, a theorem giving
sufficient conditions for $\tau(\phi)$ to depend continuously on the
holomorphic map $\phi$. The analogous result in Stuart's analysis
of slowly moving abelian Higgs vortices is Lemma 3.2 of \cite{stu1}. Our
proof of Theorem \ref{taucty}
is inspired by Stuart's argument.

\begin{thm}
\label{taucty}
Let $\phi_n:M \ra N$ be a sequence of smooth holomorphic maps between
compact
K\"ahler manifolds converging in $C^1$ to the
smooth holomorphic map $\phi:M \ra N$. Suppose all the Jacobi fields of
$\phi$ are integrable and that the conclusions of Lemma \ref{hesscty}
hold. Then
$$ \lim_{n\ra \infty}{\tau(\phi_n)} = \tau(\phi).$$
\end{thm}

\noindent {\it Proof:}
Let $P_n$, $P:H^1(E) \ra H^1(E)$ denote $L^2$ orthogonal projection onto
$\ker{\Dd^{\phi_n}}^{\perp}$ and $\ker{\Dd^{\phi}}^{\perp}$ respectively.
Since $\phi$ is Jacobi integrable, the space of sufficiently $C^1$-close
holomorphic maps is a $C^\infty$ manifold of
dimension equal to $\dim{\ker{\Dd^{\phi}}}$. This also holds for all
$\phi_n$ for all $n$ sufficiently large.
In particular, the subspaces $\ker{\Dd^{\phi_n}}$ satisfy the hypotheses of
the previous proposition for any $k$, and in particular
for $k=1$. Hence by the previous proposition we have $$||P-P_n||_1 =d_n,$$
for some positive numbers $d_n$ tending to $0$ as $n \ra \infty$.

We will prove (i) $\tau(\phi) \ge \limsup_{n\ra \infty}{\tau(\phi_n)}$
and (ii) $\tau(\phi) \le \liminf_{n\ra \infty}{\tau(\phi_n)}$. From (i) and
(ii) it
follows that $\lim_{n\ra \infty}{\tau(\phi_n)}$ exists and equals
$\tau(\phi)$.\\

Proof of (i): consider for any $V\in H^1(E) \cap
(\ker{\Dd^{\phi}})^{\perp}$, the section $V_n:=P_n V \in H^1(E)
\cap (\ker{\Dd^{\phi_n}})^{\perp}$.
Since $|| V - V_n ||_{H^1_\phi} = ||(P-P_n)V||_{H^1_\phi} \le d_n
||V||_{H^1_{\phi}}$
we get
\beq
\left| \langle V_n,V_n \rangle_{H^1_{\phi}} - \vh \right| \le
||V-V_n||_{H^1_\phi}(||V||_{H^1_\phi}+||V_n||_{H^1_\phi})
\le d_n(d_n + 2) \vh .\label{dndn}
\eeq
Using the previous inequality and inequality (ii) of Lemma \ref{hesscty} we
have
\bea
\left|{\vh} - \vhnn \right| &\le& \left| \vhnn - \langle V_n,V_n
\rangle_{H^1_{\phi}} \right| +
\left| \langle V_n,V_n \rangle_{H^1_{\phi}} - \vh \right| \nonumber \\
&\le& b_n \langle V_n,V_n \rangle_{H^1_{\phi}} +  d_n(d_n + 2) \vh
\nonumber \\
&\le& f_n \vh
\eea
where $f_n:= \left(b_n + d_n(d_n + 2)(1+b_n)\right)$ is another sequence of
positive numbers tending to zero as $n\ra \infty$.

Since $V= V_n+ K_n$ for some
$K_n\in \ker{\Dd^{\phi_n}}$, we have $\hess_{\phi_n}(V_n, V_n) =
\hess_{\phi_n}(V,V)$. Hence
\bea
\lefteqn{\left| \frac{\hp(V,V)}{\vh} - \frac{\hpn(V_n,V_n)}{\vhnn} \right|
= \left| \frac{\hp(V,V)}{\vh} - \frac{\hpn(V,V)}{\vhnn} \right|}
\nonumber \\
&\le& \frac{\hp(V,V) \left| \vh - \vhnn \right| + \vh \left| \hp(V,V) -
\hpn(V,V) \right|}{\vh \vhnn} \nonumber\\
&\le& \frac{f_n \hp(V,V) + c_n \vh}{\vhn} \le \frac{f_n}{1-f_n}
\frac{\hp(V,V)}{\vh} + \frac{c_n}{1-f_n}. \label{hessdiff}
\eea
Let $\{V^j\}_{j=1}^{\infty} \in H^1(E)\cap (\ker{\Dd^{\phi}})^{\perp}$ be a
minimizing sequence for $\tau(\phi)$ with unit $H^1_\phi$ norm, that is
$$ \lim_{j\ra \infty}{\frac{\hp(V^j,V^j)}{\langle V^j,V^j
\rangle_{H^1_{\phi}}}} = \lim_{j\ra \infty}{\hp(V^j,V^j)}= \tau(\phi).$$
Hence there exists a positive constant $C$ and numbers $\epsilon_j$ tending
to zero as $j\ra \infty$, such that
$$\hp(V^j,V^j) = \tau(\phi) + \epsilon_j$$
and
$$\hpn(V^j,V^j) \le C$$
for all $j$ and $n$. Define $V^j_n = P_n V^j \in H^1(E) \cap
(\ker{\Dd^{\phi_n}})^{\perp}$. The two previous facts together with
(\ref{hessdiff}) imply that
$$ \left| \tau(\phi) + \epsilon_j - \frac{\hpn(V^j_n,V^j_n)}{\langle V^j_n,
V^j_n \rangle_{H^1_{\phi_n}}} \right| \le \frac{Cf_n +c_n}{1-f_n}$$
holds for all $j$ and $n$. For any fixed $n$, taking the limit as $j\ra
\infty$ we see that
$$ \tau(\phi_n) \le \tau(\phi) + \frac{Cf_n +c_n}{1-f_n}.$$
Hence for any limit point $l$ of the sequence
$\{\tau(\phi_n)\}_{n=1}^{\infty}$ we have $l \le \tau(\phi)$. In particular
 we
have $$ \limsup_{n\ra \infty}{\tau(\phi_n)} \le \tau(\phi)$$
as required.\\

Proof of (ii): Since the proof of (ii) is very similar in
character to the proof of (i) we shall omit some details.
Consider the projection $PV_n$ of an element $V_n \in H^1(E) \cap
(\ker{\Dd^{\phi_n}})^{\perp}$ into $H^1(E) \cap(\ker{\Dd^{\phi}})^{\perp}$.
Several applications of the triangle inequality, together with the
inequalities
in Lemma \ref{hesscty} and the fact that $P_n\ra P$ in $H^1$, show that
there
exist two sequences of positive numbers $\{g_n\}$ and $\{h_n\}$ with
$\lim_{n\ra \infty}{g_n}=\lim_{n\ra \infty}{h_n}=0$ so that
\beq
|\vhnn - \langle PV_n,PV_n \rangle_{H^1_{\phi}}| \le g_n \langle PV_n,PV_n
\rangle_{H^1_{\phi}}
\eeq
and
\beq|\langle V_n,V_n \rangle_{H^1_{\phi}}- \langle PV_n,PV_n
\rangle_{H^1_{\phi}}| \le h_n \langle V_n,V_n
\rangle_{H^1_{\phi}}
\eeq
hold for any $V_n \in H^1(E)
\cap(\ker{\Dd^{\phi_n}})^{\perp}$. These two inequalities,
combined with the inequalities of Lemma \ref{hesscty} and the fact
that $\hp(PV_n,PV_n) = \hp(V_n,V_n)$ prove that
\bea
\lefteqn{\left| \frac{\hp(PV_n,PV_n)}{\langle PV_n,PV_n
\rangle_{H^1_{\phi}}} - \frac{\hpn(V_n,V_n)}{\vhnn} \right| =
\left| \frac{\hp(V_n,V_n)}{\langle PV_n,PV_n \rangle_{H^1_{\phi}}} -
\frac{\hpn(V_n,V_n)}{\vhnn} \right|} \nonumber \\
&\le& \frac{\hpn(V_n,V_n) |\vhnn - \langle PV_n,PV_n \rangle_{H^1_{\phi}}| +
\vhnn | \hp(V_n,V_n) - \hpn(V_n,V_n) |}
{\vhnn \langle PV_n,PV_n \rangle_{H^1_{\phi}}} \nonumber \\
&\le& \frac{g_n \hpn(V_n,V_n) \ \langle PV_n,PV_n \rangle_{H^1_{\phi}} + c_n
\vhnn \langle V_n, V_n \rangle_{H^1_{\phi}}}
{\vhnn \langle PV_n,PV_n \rangle_{H^1_{\phi}}} \nonumber \\
&\le& g_n \frac{\hpn(V_n,V_n)}{\vhnn} + c_n \frac{\langle V_n,V_n
\rangle_{H^1_{\phi}}}{\langle PV_n,PV_n \rangle_{H^1_{\phi}}}
\le g_n \frac{\hpn(V_n,V_n)}{\vhnn} + \frac{c_n}{1-h_n} \label{hessdiff2}
\eea
holds for any $V_n \in H^1(E) \cap(\ker{\Dd^{\phi_n}})^{\perp}$.

Let $\{V^j_n\}_{j=1}^{\infty} \in H^1(E)\cap (\ker{\Dd^{\phi_n}})^{\perp}$
be
a minimizing sequence for $\tau(\phi_n)$ with unit $H^1_{\phi_n}$ norm,
that is
$$ \lim_{j\ra \infty}{\frac{\hpn(V^j_n,V^j_n)}{\langle V^j_n,V^j_n
\rangle_{H^1_{\phi_n}}}} = \lim_{j\ra \infty}{\hpn(V^j_n,V^j_n)}=
\tau(\phi_n).$$
Hence there exists a positive constant $C$ and numbers $\epsilon^j_n$
tending
to zero as $j\ra \infty$, such that
$$\hp(V^j_n,V^j_n) = \tau(\phi_n) + \epsilon^j_n$$
and
$$\hp(V^j_n,V^j_n) \le C$$
for all $j$ and $n$.
Consider the sequence $\wt{V}^j_n = PV^j_n\in H^1(E) \cap
(\ker{\Dd^{\phi}})^{\perp}$.
The two previous facts together with (\ref{hessdiff2}) imply that
$$ \left| \frac{\hp(\wt{V}^j_n,\wt{V}^j_n)}{\langle \wt{V}^j_n,\wt{V}^j_n
\rangle_{H^1_{\phi}}} -  (\tau(\phi) + \epsilon^j_n) \right| \le Cg_n +
\frac{c_n}{1-h_n}$$
holds for all $j$ and $n$. For any fixed $n$, taking the limit as $j\ra
\infty$ we see that
$$ \tau(\phi) \le \tau(\phi_n) + Cg_n + \frac{c_n}{1-h_n}.$$
Hence for any limit point $l$ of the sequence
$\{\tau(\phi_n)\}_{n=1}^{\infty}$ we have $\tau(\phi)\le l$. In particular
$$ \tau(\phi) \le \liminf_{n\ra \infty}{\tau(\phi_n)} $$ as claimed.
\hfill $\Box$\\

\section{The case of $\rat_1$}
\label{rat1}
\news

 It is interesting
to consider the simplest nontrivial case, namely $\M_1=\hol_1(S^2,S^2)=
\rat_1$, in
detail. Although $M=N=S^2$ in this case, it is often helpful to distinguish
between domain and codomain by continuing to denote them $M,N$
respectively. We
shall do this when clarity requires.
The $G$ action on $\M_n$ introduced in section
\ref{sec:cp1} extends naturally to $C^\infty(S^2,S^2)$, and leaves
$E[\phi]$ invariant. It follows that $\tau(\phi)$ is constant on
$G$ orbits in $\M_n$. Since $G$ acts with cohomogeneity 1 on
$\M_1$, it suffices to consider $\tau(\phi)$ for the one-parameter
family of maps \beq \phi_\mu:z\mapsto \mu z,\qquad
\mu\in[1,\infty) \eeq lying on the curve
$\Gamma=\{([\I_2],(0,0,\lambda)):\lambda\geq 0\}$ in $\M_1$ (so
$\mu=(\sqrt{1+\lambda^2}+\lambda)^2$ as in section \ref{sec:cp1}).
We may think of $\tau$ as a
positive function $\tau(\mu)$ on $[1,\infty)$.
We will first use the results of section \ref{cty} to prove that $\tau(\mu)$
is continuous. Positivity and continuity of $\tau$ ensure that it is bounded
away from zero on any compact set.
The domain of $\tau$  (whether thought of
as a function on $\M_1$ or on $[1,\infty)$)
is noncompact, however, so we cannot conclude
that $\tau$ is globally bounded away from zero.
 The
essential question is, then, how does $\tau(\mu)$ behave as
$\mu\ra\infty$, that is, as the lump collapses to zero width? We
will prove that $\tau(\mu)\ra 0$.

First we address the issue of continuity. By symmetry,
it suffices to consider a
sequence of degree 1 holomorphic maps $\phi_n$ in the curve $\Gamma$,
labelled by a sequence $\mu_n$ in $[1,\infty)$.
If $\mu_n\ra\hat{\mu}$ then the corresponding maps $\phi_n:z\mapsto\mu_n z$
converge
in $C^1$ to $\hat{\phi}:z\mapsto\hat{\mu} z$. Since
every holomorphic map $S^2\ra S^2$ is Jacobi integrable, we may apply the
results of section \ref{cty}. We merely need to prove that the pointwise
inequality (\ref{ptwise}) on the pullback connexions $\nabla_n$
($\nabla^{\phi_n}$
transfered
to $\hat{\phi}^* TN$)
 holds for any such
sequence $\mu_n$. Then Lemma \ref{hesscty} applies, and continuity of
$\tau$ follows from Theorem \ref{taucty}.

We must first construct the canonical isometry between each  bundle
$\phi_n^*TN$ and the fixed bundle $\hat{\phi}^*TN$
equipped with their $L^2$ inner products. To this end, it is convenient
to define an orthonormal frame $E_1=\cd/\cd\theta$,
$E_2=\cosec{\theta}\ \cd/\cd\ph$ on $S^2$, where $(\theta,\ph)$ are the
usual polar coordinates, and the corresponding sections $\wt{E}_i^n
=E_i\circ\phi_n\in
\Gamma(\phi_n^*TN)$, $\hat{E}_i=E_i\circ\phi\in\Gamma(\phi^*TN)$, $i=1,2$.
Note that varying $\phi$ within the family $\Gamma$ sends each fixed $p\in M$
along a geodesic of constant $\ph$ in $N$. Note also that the frame $E_1,E_2$
is parallel along geodesics of constant $\ph$, so the canonical isometry
between $\phi^*_nTN$ and $\hat{\phi}^*TN$ is simply given by the
identification
\beq\label{ident}
\wt{E}_1^n\equiv\hat{E}_1,\quad
\wt{E}_2^n\equiv\hat{E}_2.
\eeq
It is this property which makes polar coordinates particularly natural
for our purposes.

In this coordinate system, a holomorphic map $\phi:z\mapsto \mu z$ is
\beq
(\theta,\ph)\mapsto
(f_\mu(\theta),\ph),\qquad f_\mu(\theta)=2\cot^{-1}\left(\mu \cot{
\frac{\theta}{2}}\right).
\eeq
We will construct the pullback connexion
$\nabla^\phi$ by computing its action on $\wt{E}_i=
E_i\circ\phi$, $i=1,2$.
The Levi-Civita
connexion on $S^2$ is
\beq\label{1x}
\nabla{E_1}=\cot\theta\, e_2\otimes E_2,\quad
\nabla E_2 =-\cot\theta\, e_2\otimes E_1,
\eeq
or equivalently,
\beq
\nabla e_1=\cot\theta\, e_2\otimes e_2,\quad
\nabla e_2 =-\cot\theta\, e_2\otimes e_1,
\eeq
where $e_1,e_2$ is the coframe dual to $E_1,E_2$.
The following properties
of $\nabla^\phi$ are essential for computations:
\begin{itemize}
\item[(a)] if $u\in T_pM$, $Y\in\Gamma(\phi^*TN)$ and
$g\in C^\infty(M)$, then $\nabla^\phi_ufY=u[f]Y+f\nabla^\phi_u Y$;
\item[(b)]
 if $Y\in\Gamma(\phi^*TN)$ may locally be identified with a vector field
$\wt{Y}$ on $N$ (i.e.\ $Y=\wt{Y}\circ\phi$, on a neighbourhood of $p$), then
$\nabla^\phi_u Y=\nabla^N_{d\phi u}\wt{Y}$.
\end{itemize}
Now $d\phi E_1=f_\mu'\wt{E}_1$ and $d\phi E_2=\cosec\theta\sin f_\mu
\wt{E}_2$, where $'$ denotes differentiation with respect to $\theta$. Hence
\bea
&&
\nabla^{\phi}_{E_1}\wt{E}_1=
\nabla^{\phi}_{E_1}\wt{E}_2=0
\nonumber \\
\nabla^{\phi}_{E_2}\wt{E}_1&=&\frac{\sin f_\mu}{\sin\theta}
\nabla_{\wt{E}_2}\wt{E}_1=\frac{\sin f_\mu}{\sin\theta}\cot f_\mu \wt{E}_2=
\frac{\cos f_\mu}{\sin\theta} \wt{E}_2
\nonumber \\ \label{2x}
\nabla^{\phi}_{E_2}\wt{E}_2&=&\frac{\sin f_\mu}{\sin\theta}
\nabla_{\wt{E}_2}\wt{E}_2=-\frac{\sin f_\mu}{\sin\theta}\cot f_\mu \wt{E}_2=
-\frac{\cos f_\mu}{\sin\theta} \wt{E}_2.
\eea
Note that when $\mu=1$, $\phi=\id$, and $\nabla^\id=\nabla$,
so (\ref{2x}) should reduce
to (\ref{1x}), which it does.

Any section of $\phi^*TN$ may be written
\beq
W=V+J^NU,\quad V=\Vv(\theta,\ph)\wt{E}_1,\quad
U=\Uu(\theta,\ph)\wt{E}_1.
\eeq
Note that $\nabla^\phi$ commutes with $J^N$ by the K\"ahler property. By the
defining property (a),  above, and (\ref{2x}), one sees that
\beq\label{5x}
\nabla^\phi V=\Vv_\theta\, e_1\otimes\wt{E}_1+\frac{1}{\sin\theta}\Vv_\ph\,
e_2\otimes \wt{E}_2+\Vv\frac{\cos f_\mu}{\sin\theta}\, e_2\otimes\wt{E}_2,
\eeq
where subscripts $\theta,\ph$ denote partial derivatives.
This gives us explicit formulae for $\nabla^{\hat{\phi}}$ and
$\nabla^{\phi_n}$. Using the identification (\ref{ident}) we may transfer
a section $W$
of $\hat{\phi}^*TN$ to $\phi_n^*TN$, act with $\nabla^{\phi_n}
$, then transfer back using the same identification, calling the result
$\nabla_nW$. The difference between $\nabla_nW$ and $\nabla^{\hat{\phi}}W=
\hat{\nabla}W$ is, by (\ref{5x}),
\beq\label{6x}
(\hat{\nabla}-\nabla_n)W=\frac{\cos f_{\hat{\mu}}-\cos f_{\mu_n}}{\sin\theta}
(\Vv e_2\otimes\hat{E}_2 -\Uu e_2\otimes\hat{E}_1)
=\frac{\cos f_{\hat{\mu}}-\cos f_{\mu_n}}{\sin\theta}J^NW.
\eeq
An elementary calculation shows that
\bea
\frac{\cos f_{\hat{\mu}}
-\cos f_{\mu_n}}{\sin\theta}&=&
\frac{(\hat{\mu}^2-\mu_n^2)\sin 2\theta}{2(\hat{\mu}^2\cos^2\frac{\theta}{2}+
\sin^2\frac{\theta}{2})(\mu_n^2\cos^2\frac{\theta}{2}+
\sin^2\frac{\theta}{2})}\nonumber \\
\Rightarrow\quad\left|\frac{\cos f_{\hat{\mu}}
-\cos f_{\mu_n}}{\sin\theta}\right|&\leq&\frac{1}{2}\left(1+
\frac{1}{\hat{\mu}^2}+\frac{1}{\mu_n^2}+\frac{1}{\hat{\mu}^2\mu_n^2}\right)
|\hat{\mu}^2-\mu_n^2|.\label{compcon}
\eea
Hence, we have a pointwise bound of the form (\ref{ptwise}),
\beq
|(\hat{\nabla}-\nabla_n)W|\leq a_n |W|
\eeq
where
\beq
a_n=\frac{1}{2}\left(1+
\frac{1}{\hat{\mu}^2}+\frac{1}{\mu_n^2}+\frac{1}{\hat{\mu}^2\mu_n^2}\right)
|\hat{\mu}^2-\mu_n^2|\ra 0
\eeq
as $n\ra\infty$, as required.

Having shown that $\tau(\mu)$ is continuous, we now address its behaviour
as $\mu\ra\infty$. In particular, we will
 prove that $\lim_{\mu\ra \infty}{\tau(\mu)}=0$.
To do this, it suffices to consider only sections of
$\phi^*TN$ of
a certain type, which we will call
``irrotational,'' namely those of the form
\beq
V=\Vv(\theta)\wt{E}_1.
\eeq
It is straightforward to compute $\Dd^\phi V$ for such sections,
and hence obtain
$\hess_{\phi}(V,V)$ as an explicit
integral functional of $\Vv$.  It follows from (\ref{5x}) that
\beq
\Dd^{\phi}V=-\left(\Vv'-\frac{\cos f_\mu}{\sin\theta}\Vv\right)
(e_1\otimes\wt{E}_2+e_2\otimes\wt{E}_1).
\eeq
The Hessian for irrotational sections is
\beq
\hess_{\phi}(V,V)=\half||\Dd^{\phi}V||_{L^2}^2=
2\pi\int_0^\pi d\theta\sin\theta\left(\Vv'-\frac{\cos f_\mu}{\sin\theta}\Vv
\right)^2.
\eeq
We seek to compare this quantity, for irrotational
sections $L^2$ orthogonal to $\ker\Dd^\phi$, with $||V||_{H^1}^2$.
Note that $\ker{\Dd^\phi}$ is six-dimensional, and is spanned by
\beq
K_\mu^m=-\frac{2\cot^m(\theta/2)}{1+\mu^2\cot^{2}(\theta/2)}[
\cos(1-m)\ph\, \wt{E}_1+\sin(1-m)\ph\, \wt{E}_2],\quad
m=-1,0,1
\eeq
and their images under $J^N$. This basis is obtained by considering
curves in $\M_1$ through $\phi$ generated by altering the real part
 of one of the
coefficients of the rational map.
Every irrotational section $V$ is automatically $L^2$ orthogonal to
$K_\mu^{-1},K_\mu^0,J^NK_\mu^{-1},J^NK_\mu^0$ and $J^NK_\mu^1$, so we need
only insist that $V$ is $L^2$ orthogonal to the one and only irrotational
section in the basis, $K_\mu^1$, which we will henceforth denote $K_\mu$.
Explicitly, we require that
\beq
\langle V,K_\mu \rangle_{L^2}=-2\pi\int_0^\pi d\theta\sin\theta\,
\frac{\Vv(\theta)\sin\theta}{\sin^2\frac{\theta}{2}+
\mu^2\cos^2\frac{\theta}{2}}=0,
\eeq
having rearranged $K_\mu$ slightly.

We shall also need an explict formula for $||V||_{H_1}$.
Equation (\ref{5x}) implies that
\beq
\nabla^{\phi}V=\Vv' e_1\otimes\wt{E}_1+\Vv\frac{\cos f_\mu}{\sin\theta}
e_2\otimes\wt{E}_2,
\eeq
so
\bea
||V||_{H^1}^2&=&||\nabla^{\phi}V||_{L^2}^2+||V||_{L^2}^2=
2\pi\int_0^\pi d\theta\sin\theta\left( (\Vv')^2
+\frac{\cos^2f_\mu}{\sin^2\theta}\Vv^2+\Vv^2\right)\nonumber \\
&=&\hess_{\phi}(V,V)+||V||_{L^2}^2
+2\pi\int_0^\pi d\theta\, 2\Vv\Vv'\cos f_\mu \nonumber \\
&=& \hess_{\phi}(V,V)+||V||_{L^2}^2
+\langle V,\en_\mu V\rangle_{L^2}\label{3x}
\eea
where $\en_\mu\in C^\infty(S^2)$ is the energy density of the map $\phi$.
To see the last equality, note that $\phi$ is holomorphic, hence
conformal, so
\beq\label{4x}
f_\mu'=|d\phi E_1|=|d\phi E_2|=\frac{\sin f_\mu}{\sin\theta}.
\eeq
Equation (\ref{3x}) follows from (\ref{4x}) and integration by parts.

It also proves useful to write down the formulae above using cylindrical
coordinates instead of spherical polar coordinates on the domain.
That is in place of $(\theta,\ph)$ we use $(s,\ph)$ where
$s:= \log{\cot{\theta/2}}$. First notice, that with respect to these
cylindrical coordinates $f_{\mu}(s)$ arises by translating one fixed
profile by a $\mu$ dependent amount. More precisely,
\beq
 f_{\mu}(s) = f_1(s+ \log{\mu})
\eeq
where
\beq
f_1(s) = 2 \cot^{-1}{(\exp{s})}.
\eeq
We have an analogous formula for $K_{\mu}$ the irrotational part of the
kernel of $\Dd^{\phi}$
\beq
K_{\mu}(s) = \sech{(s+ \log{\mu})} \wt{E}_1.
\eeq
It is a routine computation to obtain the following formulae for
$\hess_{\phi_{\mu}}(V,V)$, $\langle V,\en_\mu V\rangle_{L^2}$ and the
$L^2$, $H^1$ norms of an irrotational section $V$ from the corresponding
formulae using spherical polars already presented
\beq
\langle V,V \rangle_{L^2} = \int_{-\infty}^{\infty} ds\  \sech^2{s}\  \Vv^2
\eeq
\beq
\langle V,\en_\mu V\rangle_{L^2} = \int_{-\infty}^{\infty} ds\ \sech^2(s +
\log{\mu})\Vv^2 \label{evv}
\eeq
\beq
\langle V,V \rangle_{H^1} = \int_{-\infty}^{\infty} ds\
\left(\left(\frac{d\Vv}{ds}\right)^2 + \tanh^2{(s + \log{\mu})}\  \Vv^2  +
\sech^2{s}\  \Vv^2 \right)
\eeq
\beq
\hess_{\phi_{\mu}}(V,V) = \int_{-\infty}^{\infty} ds\ \left(
\frac{d\Vv}{ds} + \tanh{(s + \log{\mu}) \Vv}  \right)^2.
\eeq

We may now state and prove the main result of this section.
We shall denote the
map $z\mapsto \mu z$ by $\phi_\mu$ to emphasize its parametric dependence.

\begin{thm} [global coercivity of the Hessian fails on $\rat_1$]\label{noglo}
$\lim_{\mu \ra \infty}{\tau(\phi_{\mu})}=0.$
\end{thm}

\noindent
{\it Proof:} Define the following sequence of smooth irrotational sections
$V_{\mu}=\Vv_{\mu}\wt{E}_1$ of $\phi_\mu^*TS^2$
where
\beq
\Vv_{\mu}(s) = \sech{s} - c_{\mu} K_{\mu}(s)
\eeq
and $c_{\mu}$ is a constant determined by the requirement that
$\langle V_{\mu},K_{\mu} \rangle_{L^2}=0$.
By our previous remarks, this suffices to ensure that $\Vv_\mu$ is orthogonal
to $\ker\Dd^{\phi_\mu}.$
Since $V_{\mu} \in H^1(\phi_\mu^*TS^2) \cap \kd$ it suffices to prove that
\beq
\lim_{\mu \ra \infty} \frac{\hess_{\phi_{\mu}}(\vm,\vm)}{\langle \vm,\vm
\rangle_{H^1}^2}=0. \label{h1overhess}
\eeq
In fact, we will show that
\beq
\lim_{\mu \ra \infty} \frac{\hess_{\phi_{\mu}}(\vm,\vm)}{\langle
\vm,\en_\mu \vm\rangle_{L^2}}=0, \label{h1overev}
\eeq
which implies (\ref{h1overhess}) since
$$||V||_{H^1}^2=\hess_{\phi_\mu}(V,V)+||V||_{L^2}^2 +\langle V,\en_\mu V
\rangle_{L^2} > \langle V,\en_\mu V\rangle_{L^2}.$$

To prove (\ref{h1overev}), first notice that since $K_{\mu} \in \kd$ we have
$$\hess_{\phi_{\mu}}(\vm,\vm)= \hess_{\phi_{\mu}}(\sech{s},\sech{s}).$$
Using the explicit expression for the Hessian in cylindrical coordinates we
find
$$\hess_{\phi_{\mu}}(\sech{s},\sech{s}) = \int_{-\infty}^{\infty} ds\
\sech^2{s}\left( \tanh{\sm}-\tanh{s} \right)^2$$
where $\sm:= s + \log{\mu}$. Since $|\left( \tanh{\sm}-\tanh{s} \right)^2| <
 4$ holds for all $s$,
$$\hess_{\phi_{\mu}}(\sech{s},\sech{s}) < 4 \int_{-\infty}^{\infty} ds\
\sech^2{s} = 8$$
holds for all $\mu$. Since $\hess_{\phi_{\mu}}(\vm,\vm)$ is bounded for
all $\mu$, (\ref{h1overev}) is implied by
\beq
\lim_{\mu\ra \infty}{\langle \vm,\en_\mu \vm\rangle_{L^2}}= +\infty.
\label{evvinf}
\eeq

To establish (\ref{evvinf}), first introduce for each nonnegative pair of
integers $(m,n)$ the positive function
$$ I_{m,n}(\mu) = \intinf {\sech^m{s}\ \sech^n{s_\mu}}.$$
Clearly, the function $I_{m,n}(\mu)$ is bounded above by the constant
$I_{m,0}$.
From the definition of $\cm$ we have
$$\cm = \frac{\langle \sech,\km \rangle}{\langle \km,\km \rangle}=
\frac{I_{3,1}}{I_{2,2}}\ ,$$
and from equation (\ref{evv}) and the definition of $\vm$ we have
$$ \langle \vm,\en_\mu \vm\rangle_{L^2} = \cm \left(\cm I_{4,0} -
2 I_{1,3} \right) + I_{2,2}.$$
Since $I_{4,0}$ is a positive constant and $I_{1,3}$, $I_{2,2}$ are
positive and bounded above, (\ref{evvinf})  will follow
if we can establish that $\lim_{\mu\ra \infty}{\cm} = +\infty$. In fact,
one can explicitly evaluate the
integrals which appear in the definition of $c_{\mu}$ to obtain
$$\cm = \frac{I_{3,1}}{I_{2,2}} =
\frac{\mu^4 - 4\mu^2 \log{\mu} - 1}{4\mu( \mu^2 \log{\mu} - \mu^2 +
\log{\mu} + 1)} .$$
Clearly, this implies that $\lim_{\mu\ra \infty}{\cm} = +\infty$ as
required.\hfill
$\Box$

\section{The case of $\rat_n^{eq}$}
\label{ratn}
\news

Staying in the context of the $\CP^1$ model on $S^2$, but generalizing to
any degree sector, $n\geq 1$, things become rather more difficult. The
reason is that the isometric action of $G$ on $\M_n=\rat_n$ does not have
cohomogeneity 1, except for $n=1$. In general, then, to understand the
global behaviour of $\tau$ one must consider the $4(n-1)$ dimensional
orbifold $\M_n/G$, not simply a curve, as in section \ref{rat1}.
Reduction to a curve
 {\em is} possible, however, provided we restrict attention
to dynamics within a certain equivariance class.
Employing stereographic coordinates $z,W$ on domain and codomain as usual,
and defining polar coordinates such that $z=re^{i\ph}$, one may write the
field equation for the $\CP^1$ model on spacetime $S^2\times\R$ as
\beq
\label{el}
\frac{W_{tt}}{(1+r^2)^2}-\left(W_{rr}+\frac{1}{r}W_r+\frac{1}{r^2}W_{\ph\ph}
\right)-\frac{2\ol{W}}{1+|W|^2}\left(\frac{W_t^2}{(1+r^2)^2}-W_r^2-
\frac{1}{r^2}W_\ph^2\right)=0.
\eeq
This supports equivariant solutions within the ansatz
\beq\label{an}
W(r,\ph,t)=r^nq(r,t)e^{in\ph}.
\eeq
Provided $q:[0,\infty)\times(-\eps,\eps)\ra\C$ is nowhere vanishing
and has suitable boundary behaviour, solutions within this ansatz have degree
$n$. Substituting (\ref{an}) into (\ref{el}) one obtains a
$(1+1)$-dimensional
hyperbolic partial differential equation for $q$. The space of static
solutions is simply $q(r,t)=c\neq 0$, a complex constant. So the
equivariant moduli space is $\M_n^{eq}=\rat_n^{eq}=\{z\mapsto
 cz^n:c\neq 0\}\cong\C^\times$, the punctured
complex plane. Note that $\M_n^{eq}$ is one connected component of
the fixed point set in $\M_n$ of
the isometry group $G^{eq}=\{([\exp -in\frac{\psi}{2}\tau_3],
[\exp i\frac{\psi}{2}\tau_3]):\psi\in\R\}\cong SO(2)$, so we are assured
that $\M_n^{eq}\subset\M_n$ is totally geodesic. Both the true wave map
flow and its geodesic approximant stay within the equivariance
 class, therefore,
and one can again ask to what extent the analytic method of Stuart applies.
The equivariant configuration space enjoys a residual $U(1)$ symmetry,
namely $W(z)\mapsto e^{i\psi} W(z)$, and the discrete symmetry $W(z)\mapsto
1/W(1/z)$,
both of which leave the harmonic map energy unchanged, so we may again
restrict attention to a curve in $\M_n^{eq}$, namely
$\Gamma=\{z\mapsto \mu z^n: \mu\in [1,\infty)\}$.
The situation is actually {\em simpler} than in the full degree 1 case,
because
now the error, that is, the section of $\phi^*TN$ along which we exponentiate
to get from the holomorphic approximant to the true solution at time $t$,
must
also lie within the equivariance class. The only sections of relevance,
therefore, are irrotational sections and their $J^N$ images. We may
define a new optimal constant $\tau^{eq}(\phi)$, again the infimium
of the Hessian for $H^1$ unit sections $L^2$ orthogonal to $\ker\Dd^\phi$,
but now
we include only equivariant sections. Since these form a subset of all
sections, we have trivially that $\tau^{eq}(\phi)\geq\tau(\phi)>0$ by
Theorem \ref{coercivity}.

Once again, we may think of $\tau^{eq}$ as a function
of $\mu\in[1,\infty)$. We first show that $\tau^{eq}(\mu)$ is continuous.
Since the setup is very similar to section \ref{rat1} we
use equivalent notation and conventions, and omit several details.
In spherical polar coordinates, the map $\phi:z\mapsto \mu z^n$ is
\beq
(\theta,\ph)\mapsto(f_\mu(\theta),n\ph),
\qquad f_\mu(\theta)=2\cot^{-1}\left(\mu \cot^n{
\frac{\theta}{2}}\right),
\eeq
so $d\phi\, E_1=f_\mu'\wt{E}_1$ and $d\phi\, E_2=n\, \cosec\theta\cos f_\mu
\wt{E}_2$. It follows that
\beq
\nabla^\phi\wt{E}_1=n\frac{\cos f_\mu}{\sin\theta}e_2\otimes \wt{E_2},\quad
\nabla^\phi\wt{E}_2=-n\frac{\cos f_\mu}{\sin\theta}e_2\otimes\wt{E_1}.
\eeq
Hence, on an irrotational section $V=\Vv(\theta)\wt{E}_1$,
\beq
\nabla^\phi V=V'e_1\otimes\wt{E}_1+Vn\frac{\cos f_\mu}{\sin\theta}
e_2\otimes\wt{E}_2.
\eeq
Given a general equivariant section $W=V+J^NU$, where $V,U$ are irrotational,
we may transfer $W$ to a neighbouring bundle $\hat{\phi}^*TN$ (where
$\hat{\phi}:z\mapsto\hat{\mu} z$) using the canonical isometry (\ref{ident}),
act with $\nabla^{\wh{\phi}}$, then transfer back again, to obtain
$\hat{\nabla}W$. The difference between this and $\nabla^\phi W$ is
\beq
(\nabla^\phi-\hat{\nabla})W=n\frac{\cos f_\mu-\cos f_{\hat{\mu}}}{\sin\theta}
J^NW.
\eeq
Now
\bea
\alpha(\theta)=\frac{\cos f_\mu-\cos f_{\hat{\mu}}}{\sin\theta}&=&
\frac{(\mu^2-\hat{\mu}^2)(\cos(\theta/2)\sin(\theta/2))^{2n-1}}{
(\sin^{2n}(\theta/2)+\mu^2\cos^{2n}(\theta/2))
(\sin^{2n}(\theta/2)+\hat{\mu}^2\cos^{2n}(\theta/2))}\nonumber
\\
\Rightarrow\quad |\alpha(\theta)|&\leq&
\left(1+
\frac{1}{\hat{\mu}^2}+\frac{1}{\mu^2}+\frac{1}{\hat{\mu}^2\mu^2}\right)
|\hat{\mu}^2-\mu^2|.
\eea
So the pullback connexion satisfies a pointwise bound of the correct type
($|(\nabla^\phi-\hat{\nabla})W|<c(\mu,\hat{\mu})|W|$ where $c\ra 0$ as
$\mu\ra\hat{\mu}$) and we may conclude from the results of section \ref{cty}
that $\tau^{eq}(\mu)$ is continuous.

As for $\rat_1$, the interesting issue is the large $\mu$ behaviour of
$\tau^{eq}(\mu)$. We are so far unable to prove anything rigorous about
$\lim_{\mu\ra\infty}\tau^{eq}(\mu)$ (if, indeed, it exists). However, we
make the following conjecture:

\begin{conj} For all $n\geq 1$, $\lim_{\mu\ra\infty}\tau^{eq}(\mu)$
exists and is finite. For all $n>1$, this limit is not zero.
\end{conj}

\noindent
If true, the equivariant Hessian is globally coercive for degree $n$
greater than
unity. One could at least hope, therefore, to model equivariant lump collapse
accurately within the geodesic approximation in the case $n>1$.

To motivate this conjecture, we should describe some numerical work which led
directly to the proof of its $n=1$ counterpart, Theorem \ref{noglo}. It is
straightforward to write the Jacobi operator on irrotational sections as an
explicit ordinary differential operator acting on $\Vv(\theta)$. The
eigenvalue problem for $\jac^\phi$ then reduces to a singular
Sturm-Liouville problem on $(0,\pi)$, namely
\beq
\label{sturm}
-\Vv''-\cot\theta\Vv'+n^2\frac{\cos 2 f_\mu}{\sin^2\theta}\Vv=\omega^2\Vv.
\eeq
Here $\omega^2\geq 0$ is the eigenvalue.
The eigenfunctions of this problem form an $L^2$ orthogonal basis for the
space of irrotational sections. In particular, the eigenfunctions outside the
kernel, which we shall refer to as ``excited states'', in analogy with
quantum mechanics,
 form a basis for the irrotational sections orthogonal to
$\ker\jac^\phi$. If we could prove that $\hess_\phi(V,V)/||V||_{H^1}^2=
\omega^2||V||_{L^2}^2/||V||_{H^1}^2$ is bounded uniformly away from zero for
all $\mu$
and all excited states $V$, global coercivity of $\hess^{eq}$ would follow.
Conversely, if for one of the excited states, $\omega^2 ||V||_{L^2}^2/
||V||_{H^1}^2\ra 0$ as $\mu\ra \infty$, global coercivity must fail: the
one parameter family of excited states itself provides a counterexample.
Unfortunately, the ($\mu$ families) of eigensections are rather inaccessible
analytically (except in the special case $\mu=n=1$, where they are known
exactly). However, they are quite easy to construct numerically, either by
employing a specialist Sturm-Liouville solver package, or by using a
shooting method. We have tried both strategies, obtaining compatible
results from each.

\begin{figure}
\begin{center}
\epsfysize=6truecm
\epsfbox{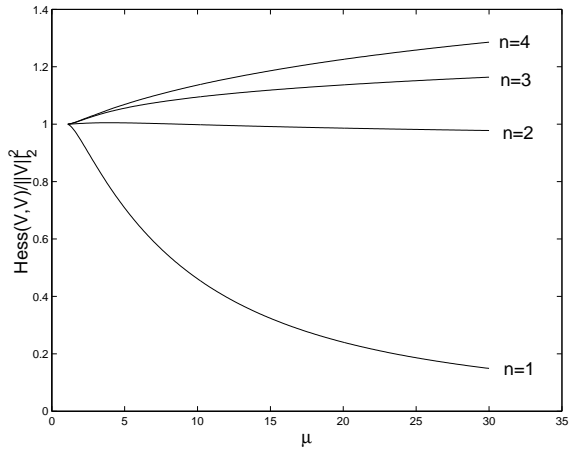}
\end{center}
\caption{\sf
Plots of the ratio $\hess_{\phi_\mu}(V_\mu,V_\mu)/||V_\mu||_{H^1}^2$ against
$\mu$ for the first excited state $V_\mu$ of the Sturm-Liouville problem
(\ref{sturm}) with $\phi_\mu:z\ra\mu z^n$, $n=1,2,3,4$. To facilitate
comparison of the curves, in each case the ratio has been normalized by its
value at $\mu=1$ ($0.30$, $0.74$, $1.17$ and $1.55$ respectively, to 2
decimal places). Note that only for $n=1$ does the ratio tend to 0 as
$\mu\ra\infty$.
}
\end{figure}

The pertinent results may be summarized as follows.
For $n=1,2,3,4$ and for all excited states from 2nd to 6th (ordered by
increasing $\omega^2$), the ratio $\hess_\phi(V,V)/||V||_{H_1}^2$ appears
to remain bounded away from zero as $\mu\ra\infty$. More interesting is the
first excited state. Here the ratio remains bounded away from zero for
$n=2,3,4$, but not for $n=1$. It was by examining the
graphs of the $n=1$
first excited eigenstates that the explicit family $V_\mu$ of the proof of
Theorem \ref{noglo} was devised: the family is designed to have the same
qualitative behaviour as the numerically generated eigenstates.
In figure 1 we present a graph showing the ratio
$\hess_\phi(V,V)/||V||_{H_1}^2$ as a function of $\mu$ for the first excited
states for $n=1,2,3,4$.
These data were generated by a shooting method using a 4th order Runge-Kutta
scheme with variable $\theta$
step. The singularities at $\theta=0,\pi$ were handled
by series expansions, so the scheme shot forwards from $\theta=\delta$ and
backwards from $\theta=\pi-\delta$, applying a matching condition at $\theta=
\frac{\pi}{2}$ ($\delta$ being a small positive number, typically $0.001$).
The difference between $n=1$ and the other cases is
quite clear. Although it is impossible to be exhaustive numerically, the
results suggest that $\hess_\phi(V,V)/||V||_{H_1}^2$ is bounded away from
zero for $n>1$, as we have conjectured.
Linhart and Sadun in a recent numerical study
of lump collapse in the
$\CP^1$ model on $\C$ imposed the analogous equivariance
condition on their field equation \cite{linsad}. It is an intriguing
coincidence that they found that single lump collapse differs greatly from
the predictions of the geodesic approximation (truncated to a finite disk in
$\C$), whereas the collapse of two coincident lumps is quite well modelled by
the geodesic flow.


\end{document}